\documentclass[journal]{IEEEtran}
\renewcommand\IEEEkeywordsname{Index Terms}
\usepackage{blindtext}
\usepackage{graphicx}
\usepackage{amssymb}
\usepackage{amsmath}
\usepackage{amsfonts}
\usepackage{xcolor}
\usepackage{braket}
\usepackage{multirow}
\usepackage{subfigure}
\usepackage{cite}
\usepackage[ruled]{algorithm2e}
\usepackage[utf8]{inputenc}
\usepackage[english]{babel}
\usepackage{caption}
\usepackage{amsthm}
\ifCLASSINFOpdf
\else
\fi
\newcommand{\va}{{\bf a}}

\newcommand{\ve}{{\bf e}}

\newcommand{\vg}{{\bf g}}

\newcommand{\vq}{{\bf q}}
\newcommand{\vr}{{\bf r}}

\newcommand{\vu}{{\bf u}}
\newcommand{\vv}{{\bf v}}

\newcommand{\vx}{{\bf x}}
\newcommand{\vy}{{\bf y}}
\newcommand{\vz}{{\bf z}}

\newcommand{\cA}{{\cal A}}
\newcommand{\cB}{{\cal B}}

\newcommand{\cI}{{\cal I}}

\newcommand{\cK}{{\cal K}}

\newcommand{\cN}{{\cal N}}

\newcommand{\cS}{{\cal S}}
\newcommand{\cT}{{\cal T}}

\newcommand{\bR}{{\mathbb{R}}}

\newcommand{\bN}{{\mathbb{N}}}

\newenvironment{mat}[1]{\left[\begin{array}{#1}}{\end{array}\right]}

\newcommand{\ceil}[1]{\left\lceil#1\right\rceil}
\newcommand{\floor}[1]{\left\lfloor#1\right\rfloor}

\renewcommand{\tt}[2]{{\cT_{#1}^{#2}}}

\newcommand{\mA}{{\bf A}}

\newcommand{\mI}{{\bf I}}

\newcommand{\mU}{{\bf U}}

\newcommand{\ignore}[1]{}

\newcommand{\modd}[1]{\left\langle#1\right\rangle}

\begin{document}
%
\newtheorem{corol}[theorem]{Corollary}
\title{Greedy Algorithms for Hybrid Compressed Sensing}
%
%
%

\author{
Ching-Lun Tai, Sung-Hsien Hsieh, and~Chun-Shien Lu 
\thanks{The authors are with the Institute of Information Science, Academia Sinica, Taipei, Taiwan (e-mail: vincent001217@gmail.com; parvaty316@hotmail.com; lcs@iis.sinica.edu.tw). }
}

\maketitle

\begin{abstract}
Compressed sensing (CS) is a technique which uses fewer measurements than dictated by the Nyquist sampling theorem.
The traditional CS with linear measurements achieves efficient recovery performances, but it suffers from the large bit consumption due to the huge storage occupied by those measurements.
Then, the one-bit CS with binary measurements is proposed and saves the bit budget, but it is infeasible when the energy information of signals is not available as a prior knowledge.
Subsequently, the hybrid CS which combines the traditional CS and one-bit CS appears, striking a balance between the pros and cons of both types of CS.
Considering the fact that the one-bit CS is optimal for the direction estimation of signals under noise with a fixed bit budget and that the traditional CS is able to provide residue information and estimated signals, we focus on the design of greedy algorithms, which consist of the main steps of support detection and recovered signal update, for the hybrid CS in this paper.
We first propose a theorem on the random uniform tessellations for sparse signals to further investigate the properties of one-bit CS.
Afterwards, we propose two greedy algorithms for the hybrid CS, with the one-bit CS responsible for support detection and traditional CS offering updated residues and signal estimates.
For each of the proposed algorithms, we provide the corresponding theorem with proof to analyze their capabilities theoretically.
Simulation results have demonstrated the efficacy of the proposed greedy algorithms under a limited bit budget in noisy environments.
\end{abstract}


%
\IEEEpeerreviewmaketitle

\section{Introduction}
With the emergence of the big data world, the amount of information available has dramatically increased during recent years.
For practical devices in signal processing and efficient algorithms in machine learning, it is still an open challenge to recover high-dimensional data with a small number of measurements.

Compressed sensing (CS) \cite{candes06,donoho06} is a technique which uses a smaller measurement rate than the Nyquist rate to recover the high-dimensional signals which have sparsity.
CS is widely adopted in various applications, such as medical imaging \cite{lustig08}, video coding \cite{baraniuk17a}, radar imaging \cite{potter10}, cognitive radio communications \cite{sharma16}, and so on.

In the traditional CS, one would obtain $m$ linear measurements of the form
\begin{equation}
    [\vy]_i=\modd{\va_i,\vx}, i=1,2,...,m,
    \label{eq:real_CS}
\end{equation}
where $\va_1,\va_2,...,\va_m\in\bR^n$ are the $m$ known vectors and $\vx\in\bR^n$ is the signal (which has sparsity) to be recovered. Note that (\ref{eq:real_CS}) can be rewritten in the form as $\vy=\mA\vx$, where the measurement vector $\vy\in\bR^m$ contains the $m$ entries $[\vy]_1,[\vy]_2,...,[\vy]_m$ and $\mA\in\bR^{m\times n}$ is the measurement matrix with the rows $\va_1,\va_2,...,\va_m$.

There are two kinds of sparse signals, $s$-sparse signals and effectively $s$-sparse signals.
For a signal $\vx\in\bR^n$, it is called $s$-sparse if
\begin{equation}
    \|\vx\|_0:=|\mbox{supp}(\vx)|=s\ll n,\nonumber
\end{equation}
or it is called effectively $s$-sparse if
\begin{equation}
    \|\vx\|_1\leq\sqrt{s}\|\vx\|_2, s\ll n.\nonumber
\end{equation}
It has been shown that (see the references in \cite{eldar12,foucart13}) when $\mA$ consists of independent standard normal entries (scaled by any nonzero real constant), one can recover any $s$-sparse signal $\vx$ from the $m\approx s\mbox{log}(n/s)$ linear measurements $[\vy]_i,i=1,2,...,m$, as defined in (\ref{eq:real_CS}) with high probability.


Despite its efficient recovery performances, the traditional CS suffers from large bit consumption due to the fact that the linear measurements are stored as numbers of high-precision data types (\emph{e.g.}, float numbers) in most systems.
In order to deal with the storage issue, the one-bit CS, which takes binary measurements, is proposed \cite{boufounos08}.

In the one-bit CS, one would obtain $m$ binary measurements of the form
\begin{equation}
    [\vy]_i=\mbox{sign}(\modd{\va_i,\vx}), i=1,2,...,m,
    \label{eq:onebit_CS}
\end{equation}
where $[\vy]_1,[\vy]_2,...,[\vy]_m\in\{\pm 1\}$, $\va_1,\va_2,...,\va_m\in\bR^n$ are the $m$ known vectors, and $\vx\in\bR^n$ is the signal (which has sparsity) to be recovered.
Note that (\ref{eq:onebit_CS}) can be rewritten in the form as
\begin{equation}
    [\vy]_i\cdot\modd{\va_i,\vx}\geq 0, i=1,2,...,m,
    \label{eq:1bit}
\end{equation}
and in the form as $\vy=\mbox{sign}(\mA\vx)$, where the measurement vector $\vy\in\{\pm 1\}^m$ contains the $m$ entries $[\vy]_1,[\vy]_2,...,[\vy]_m$ and $\mA\in\bR^{m\times n}$ is the measurement matrix with the rows $\va_1,\va_2,...,\va_m$.


Despite its less bit consumption, the one-bit CS lacks the energy information of signals due to the inherent sign operation.
Therefore, the one-bit CS would be infeasible for signal recovery when there is no prior knowledge of signal energy.

Combining both the traditional CS and one-bit CS, the hybrid CS appears in order to strike a balance between the advantages and disadvantages between both types of CS.
In our settings of the hybrid CS, one would obtain noisy linear measurements of the form
\begin{equation}
    \vy_r=\mA_r\tilde{\vx}=\mA_r\vx+\ve_r\in\bR^{m_r},
    \label{eq:linear_measurements}
\end{equation}
and noisy binary measurements of the form
\begin{equation}
    \vy_o=\mbox{sign}(\mA_o\tilde{\vx})=\mbox{sign}(\mA_o\vx+\ve_o)\in\{\pm 1\}^{m_o},
    \label{eq:binary_measurements}
\end{equation}
where
\begin{equation}
    \tilde{\vx}=\vx+\vu\in\bR^n
\end{equation}
is the contaminated signal with $\vu\in\bR^n$ being the signal-level noise,
$\mA_r\in\bR^{m_r\times n}$ and $\mA_o\in\bR^{m_o\times n}$ are the measurement matrices of the traditional CS and one-bit CS, respectively, and $\ve_r=\mA_r\vu\in\bR^{m_r}$ and $\ve_o=\mA_o\vu\in\bR^{m_o}$ are the error vectors of the traditional CS and one-bit CS, respectively.


According to the fact that the one-bit CS is optimal for estimating the directions of signals under noise with a fixed bit budget \cite{slawski18} and that the traditional CS can be used to compute residues and signal estimates, it is reasonable to have greedy algorithms, which consist of the main steps of support detection and recovered signal update, for the hybrid CS.

In this study, we first investigate several properties of one-bit CS. Then, we propose two greedy algorithms with proofs for the hybrid CS.

\subsection{Related Works}
Since the hybrid CS is established based on both the traditional CS and one-bit CS, we briefly review the recovery algorithms in both types of CS.
In addition, we survey the recent development of the hybrid CS.

For the traditional CS, the recovery algorithms can be classified into three categories: convex relaxation, non-convex optimization, and greedy algorithms \cite{carmi14}.
First, the convex relaxation algorithms (\emph{e.g.}, \cite{candes06a,candes07,Daubechies08,beck09,Donoho09,garg09}) relax the original problem into convex ones, whose efficient solutions can be derived with existing methods.
Next, the non-convex optimization algorithms (\emph{e.g.}, \cite{chartrand08,ji08}) take advantage of the knowledge of the distribution of the signal to be recovered and provide the statistics of their estimate; however, they are not suitable for the recovery of rather high-dimension signals due to the intractable computational requirements.
Lastly, the greedy algorithms (\emph{e.g.}, \cite{tropp07,blumensath08,Blumensath08a,dai09,NEEDELL09,needell09a,eldar10,donoho12,wang12,kwon14}) recover the signal iteratively, choosing a local optimum in each iteration until convergence (or when the stopping criterion is satisfied).
A detailed review of these algorithms can be found in several summary papers, \emph{e.g.}, \cite{rani18,marques19}.

For the one-bit CS, the recovery algorithms can be classified into three categories: optimization algorithms, Bayesian algorithms, and greedy algorithms.
First, the optimization algorithms (\emph{e.g.}, \cite{laska11,yan12,movahed12,kamilov12,Plan13,jacques13,plan13b,zhang14,chen15,baraniuk17}) reformulate the original problem as optimization ones, which can be solved with existing methods efficiently.
Next, the Bayesian algorithms (\emph{e.g.}, \cite{shen13,li15}) adopt the prior knowledge of the signal and provide the estimate with statistical methods.
Lastly, the greedy algorithms (\emph{e.g.}, \cite{boufounos09,liu16}) are executed in an iterative manner, recovering the signal by selecting a local optimum in each iteration until convergence (or when the stop criterion is met).

Integrating both types of CS, the hybrid CS is a rather new topic.
To our best knowledge, we only find one work \cite{huang17} which applies the hybrid CS to the overexposure problem in computed tomography.
In \cite{huang17}, the linear measurements whose magnitudes are below the threshold value $s$ will be truncated to the binary values. To solve this hybrid CS problem, the authors propose an optimization algorithm which minimizes the total variation (TV) norm of the estimated signal and the penalty terms. This approach is far different from the greedy algorithms proposed in this study.

\subsection{Contributions}
Focusing on the hybrid CS, this study provides the following three main contributions.
\begin{itemize}
    \item \textbf{Random uniform tessellations for sparse signals}: The one-bit CS is related to the field of random uniform tessellations (see Sec. \ref{subsec:RUT}). 
    Therefore, we propose a fundamental theorem for an analysis of the random uniform tessellations for sparse signals in order to demonstrate the robustness to noise and the properties of error bound of one-bit CS.
    \item \textbf{Greedy algorithms for hybrid CS}: For the hybrid CS, we propose two greedy algorithms, where the one-bit CS is responsible for support detection, and the traditional CS provides the updated residues and signal estimates.
    The first algorithm detects the support of signals, while the second algorithm modifies an initial guess of the support of signals into a more correct one.
    \item \textbf{Theoretical analysis of the proposed algorithms}: For each of the two proposed algorithms, we offer the corresponding theorem with proof to theoretically analyze its capabilities.
    For the first algorithm, we provide a theorem on the probability of successful detection of the support of signals; for the second algorithm, we provide a theorem on the probability of successful modifications of the initial guess of the support of signals into the correct one. 
\end{itemize}

\subsection{Organizations}
In Sec. \ref{sec:preliminaries}, we review several essential concepts, such as basic properties of random matrices, Gaussian mean width, and random uniform tessellations.
Then, we investigate the properties of random uniform tessellations for sparse signals with the proposed fundamental theorem in Sec. \ref{sec:main}.
In Sec. \ref{sec:method}, we introduce the proposed two greedy algorithms for the hybrid CS with corresponding theorems for an analysis of their capabilities.
Simulation results and further discussions are provided in Sec. \ref{sec:simulation}.
Finally, we conclude this study in Sec. \ref{sec:conclusion}.

\section{Preliminaries}
\label{sec:preliminaries}
In this section, we summary the notations used in this study and review several fundamental concepts related to the proposed algorithms, including basic properties of random matrices, Gaussian mean width, and random uniform tessellations.
\subsection{Notations}
The notations used in this study are summarized as follows:
Boldfaced capital and lowercase letters denote matrices and column vectors, respectively. 
For any set $\cA$, we use $|\cA|$ to denote its cardinality.
Given a vector $\vv$, we use $[\vv]_n$ to denote the $n$th component of $\vv$, $[\vv]_{\cI}$ the subvector of $\vv$ consisting of the entries indexed by the set $\cI$, $[\vv]^r$ the signal formed by restricting $\vv$ to its $r$ largest-magnitude components with the remaining entries set as zero, $\|\vv\|_2$ the $\ell_2$-norm of $\vv$, $\|\vv\|_1$ the $\ell_1$-norm of $\vv$, $\|\vv\|_0$ the number of nonzero entries of $\vv$.
Given two vectors $\vv_1$ and $\vv_2$, we use $\theta(\vv_1,\vv_2)$ to denote the angle between $\vv_1$ and $\vv_2$.
Given a matrix $\mU$, we use $\mU_{\cI}$ to denote the submatrix of $\mU$ consisting of the columns indexed by the set $\cI$, $\mU_{\{\cI:0\}}$ the matrix formed by restricting $\mU$ to the columns indexed by the set $\cI^c$ with the remaining columns set as zero vectors, $\mU^*$ the transpose of $\mU$, $\mU^{\dagger}$ the pseudo inverse of $\mU$.
We define $\cS^{n-1}:=\{\vv\in\bR^n\mid\|\vv\|_2=1\}$ to be the unit Euclidean sphere in $\bR^n$, and $\cB_1^n:=\{\vv\in\bR^n\mid\|\vv\|_1\leq 1\}$ and $\cB_2^n:=\{\vv\in\bR^n\mid\|\vv\|_2\leq 1\}$ the unit $\ell_1$-ball and unit $\ell_2$-ball in $\bR^n$, respectively.
The set $\Sigma_{n,s}:=\{\vv\in\bR^n\mid\|\vv\|_0\leq s\}$ of $s$-sparse vectors in $\bR^n$ is followed by the set $\Sigma'_{n,s}=\{\vv\in\bR^n\mid\|\vv\|_0\leq s, \|\vv\|_2\leq 1\}$ of $s$-sparse vectors in $\cB_2^n$, while the set $\cK_{n,s}=\{\vv\in\bR^n\mid\|\vv\|_1\leq \sqrt{s}\|\vv\|_2\}$ of the effectively $s$-sparse vectors in $\bR^n$ is followed by the set $\cK'_{n,s}=\{\vv\in\bR^n:\|\vv\|_1\leq \sqrt{s}, \|\vv\|_2\leq 1\}$ of the effectively $s$-sparse vectors in $\cB_2^n$.
We define ${\mI}_q$ to be the $q \times q$ identity matrix, and $\tau_n(\vv,\cI)$ the vector formed by setting the entries indexed by the set $\cI$ as $\vv$ and the remaining entries as zero.
Given a random variable $X$ which follows the binomial distribution $B(n',p^*)$, its cumulative distribution function $\mathbb{P}\{X\leq k\}=I_{1-p^*}(n'-k,k+1)=(n'-k)\binom{n'}{k}\int_0^{1-p^*}t^{n'-k-1}(1-t)^k dt$ \cite{wadsworth60}.
\subsection{Basic Properties of Random Matrices}
In this study, the measurement matrices $\mA_r\in\bR^{m_r\times n}$ and $\mA_o\in\bR^{m_o\times n}$ have independent standard normal entries which are divided by $\sqrt{m_r}$ and $\sqrt{m_o}$, respectively. The following theorems and lemmas introduce several fundamental properties of this kind of random matrix (which will be denoted as $\mA\in\bR^{m\times n}$ in this subsection). Note that the below $C$ and $c$ are positive constants.

First, we will take a look at the restricted isometry property (RIP), which is one of the most important properties developed in CS.


\textit{Definition 1 (RIP of order $s$ \cite{candes05}): A matrix $\mA$ is said to satisfy the RIP of order $s$ if there exists a constant $\delta_s\in(0,1)$ such that the following holds for any $\vx\in\Sigma_{n,s}$:
\begin{equation}
    (1-\delta_s)\|\vx\|_2^2\leq\|\mA\vx\|_2^2\leq(1+\delta_s)\|\vx\|_2^2.
\end{equation}
The constant $\delta_s$ is called the restricted isometry constant of order $s$.
}

\textit{Theorem 1 (Sufficient conditions for RIP of order $s$ \cite{foucart13}): 
Let $0<\delta_s<1$. Suppose that $m\geq C\delta^{-2}s\mbox{log}(n/s)$, then with failure probability at most $2\mbox{exp}(-c\delta^2m)$ over the choice of $\mA$, $\mA$ satisfies the RIP of order $s$.
}

With the RIP, several properties can be observed through the following lemmas.

\textit{Lemma 1 (Monotonicity of RIP \cite{candes05}): If a sensing matrix $\mA$ satisfies the RIP of both orders $s_1$ and $s_2$, then $\delta_{s_1}\leq\delta_{s_2}$ for any $s1\leq s2$.
}

\textit{Lemma 2 (Consequences of RIP \cite{candes05}): Let $\cI\subset\{1,2,...,n\}$. If a sensing matrix $\mA$ satisfies the RIP of order $|\cI|$, then for any $\vx\in\bR^{|\cI|}$,
\begin{align}
        (1-\delta_{|\cI|})\|\vx\|_2 & \leq \|\mA_{\cI}^*\mA_{\cI}\vx\|_2\leq(1+\delta_{|\cI|})\|\vx\|_2,\\
        \frac{1}{1+\delta_{|\cI|}}\|\vx\|_2 & \leq\|(\mA^*_{\cI}\mA_{\cI})^{-1}\vx\|_2\leq\frac{1}{1-\delta_{|\cI|}}\|\vx\|_2.
    \end{align}
}

\textit{Lemma 3 (Applications of RIP \cite{candes08}): Let $\cI_1,\cI_2\subset\{1,2,...,n\}$ be two disjoint sets ($\cI_1\cap\cI_2=\varnothing$). If a sensing matrix $\mA$ satisfies the RIP of order $|\cI_1|+|\cI_2|$, then for any $\vx\in\bR^{|\cI_2|}$,
 \begin{equation}
        \|\mA_{\cI_1}^*\mA_{\cI_2}\vx\|_2\leq\delta_{|\cI_1|+|\cI_2|}\|\vx\|_2.
    \end{equation}
}

Next, we will turn our attention to the $\ell_1$-quotient property, which helps with the noise analysis in the CS applications.

\textit{Theorem 2 ($\ell_1$-quotient property \cite{Wojtaszczyk10}): If $n\geq 2m$, then with failure probability at most $\mbox{exp}(-cm)$, there exists an absolute constant $d$ such that every $\ve\in\bR^m$ can be expressed as
    \begin{equation}
    \ve=\mA\vu\quad \mbox{with} \quad \|\vu\|_1\leq d\sqrt{m/\mbox{log}(n/m)}\|\ve\|_2.
    \end{equation}
}

Combining Theorem 1 and Theorem 2, the following corollary can be derived.

\textit{Corollary 1 (Simultaneous $(\ell_2,\ell_1)$-quotient property \cite{baraniuk17}): Suppose that $m\geq C\delta^{-2}s\mbox{log}(n/s)$ and $n\geq 2m$, then with failure probability at most $\mbox{exp}(-cm)$, there exist absolute constants $d$, $d'>0$ such that every $\ve\in\bR^m$ can be expressed as
    \begin{equation}
        \ve=\mA\vu \quad \mbox{with}
\left\{
\begin{array}{l}
\|\vu\|_2 \leq d\|\ve\|_2\\
\|\vu\|_1 \leq d'\sqrt{m/\mbox{log}(n/m)}\|\ve\|_2.
\end{array}
\right .
    \end{equation}
}

\subsection{Gaussian Mean Width}
Now, we will have a brief review on the Gaussian mean width. Note that the below $C$ and $C'$ are positive constants.

Consider a bounded subset $\cK\subset\bR^n$, the \emph{Gaussian mean width} of $\cK$ is defined as
\begin{equation}
    w(\cK):=\mathbb{E}\underset{\vq\in\cK}{\mbox{sup}}\modd{\vg,\vq},
\end{equation}
where $\vg\sim N(0,\mI_n)$ is a standard normal random vector in $\bR^n$.

The Gaussian mean width of several bounded subsets has been studied.
For instance, the Gaussian mean width of $\Sigma'_{n,s}$ satisfies \cite{Plan13}
\begin{equation}
    w^2(\Sigma'_{n,s})\leq Cs\mbox{log}(n/s).
\end{equation}
Since $\cK'_{n,s}$ is a convexification of $\Sigma'_{n,s}$ \cite{plan13b}, \emph{i.e.}, 
\begin{equation}
    \mbox{conv}(\Sigma'_{n,s})\subset\cK'_{n,s}\subset 2\mbox{conv}(\Sigma'_{n,s}),
\end{equation}
the Gaussian mean width of $\cK'_{n,s}$ satisfies
\begin{equation}
    w^2(\cK'_{n,s})\leq C' s\mbox{log}(n/s).
\end{equation}
Consider the bounded subset $\sqrt{s}\cB^n_1\cap\cS^{n-1}\subset\cK'_{n,s}$. Its Gaussian mean width satisfies
\begin{equation}
    w^2(\sqrt{s}\cB^n_1\cap\cS^{n-1})\leq w^2(\cK'_{n,s})\leq C' s\mbox{log}(n/s).
    \label{eq:GMW_import}
\end{equation}

\subsection{Random Uniform Tessellations}
\label{subsec:RUT}

Now, we would like to further investigate the one-bit CS which is expressed in the form as (\ref{eq:1bit}).
Note that $\modd{\va_i,\vz}=0,\forall\vz\in\bR^n,i=1,2,...,m_o$, defines a hyperplane perpendicular to $\va_i$ in $\bR^n$.
Suppose that $\vx\in\cK$, where $\cK$ is a subset of $\bR^n$ and can be considered as an oddly shaped object.
Assume that there exists another signal $\vy\in\cK$.
For a single hyperplane whose normal vector is $\va_i, i=1,2,...,m_o$, the probability of this hyperplane separating $\vx$ and $\vy$ can be expressed in terms of the angle $\theta(\vx,\vy)$ by the following lemma.\\

\textit{Lemma 4 (Probability of sign change \cite{goemans95}): Let $\vx,\vy\in\bR^n$ and $\va\sim\cN(0,\sigma^2\mI_n)$. Then,}
    \begin{equation}
        \mathbb{P}\{\mbox{sign}(\modd{\va,\vx})\neq\mbox{sign}(\modd{\va,\vy})\}=\frac{\theta(\vx,\vy)}{\pi}.
    \end{equation}


Before further discussions, we will first introduce the following definition.

\textit{Definition 2 (Fraction of hyperplanes that separate points \cite{Plan14}): For $\vx,\vy\in\bR^n$, the fraction of $m$ hyperplanes generated from $\mA\in\bR^{m\times n}$ that separate points $\vx$ and $\vy$ is denoted as}
\begin{equation}
    d_{\mA}(\vx,\vy):=|\{i|\mbox{sign}(\va_i,\vx)\neq\mbox{sign}(\va_i,\vy), i=1,2,...,m\}|/m,
\end{equation}
\textit{where $\va_i$ is the $i$th row of $\mA$.}

According to Definition 2, if $d_{\mA}(\vx,\vy)$ is smaller, 
then it can be interpreted as that the object (\emph{i.e.}, $\cK$) is sliced by more hyperplanes, which narrow down the minimum space that contains both $\vx$ and $\vy$ and the bound of the $\ell_2$-norm distance between $\vx$ and $\vy$.

The above discussions are linked to the research of random uniform tessellations, which is broadly addressed in \cite{Plan14}.
The following theorem provides a connection between $d_{\mA}(\vx,\vy)$ and the upper bound of the $\ell_2$-norm distance $\|\vx-\vy\|_2$ in the noiseless case. Note that the lower bound of $m$ is relevant to the $w^2(\cK)$.

\textit{Theorem 3 (Cells of random uniform tessellations \cite{Plan14}): Consider a bounded subset $\cK\subseteq\cS^{n-1}$ and a matrix $\mA\in\bR^{m\times n}$, which has independent standard normal entries divided by $\sqrt{m}$. 
Let $0<\delta<1$. Suppose that
\begin{equation}
    m\geq C\delta^{-4}w^2(\cK).
    \label{eq:m}
\end{equation}
Then, with probability at least $1-2\mbox{exp}(-c\delta^4 m)$ over the choice of $\mA$, the following satisfies for any $\vx,\vy\in\cK$:
\begin{equation}
    \|\vx-\vy\|_2\leq \delta+d_{\mA}(\vx,\vy).
\end{equation}
The above $C$ and $c$ are positive constants.
}

From Theorem 3, it can be observed that the upper bound of the $\ell_2$-norm distance $\|\vx-\vy\|_2$ is proportional to $d_{\mA}(\vx,\vy)$.
Therefore, we can have a smaller $\ell_2$-norm distance $\|\vx-\vy\|_2$ by decreasing $d_{\mA}(\vx,\vy)$.

\section{Random Uniform Tessellations for Sparse Signals}
\label{sec:main}
According to Sec. \ref{subsec:RUT}, the one-bit CS is closely related to the field of random uniform tessellations. In this section, we would like to further investigate the properties of random uniform tessellations in order to have a better understanding of the one-bit CS.
Therefore, we develop a fundamental theorem of random uniform tessellations on the error bound for sparse signals, extended from Sec. \ref{subsec:RUT}, to demonstrate the robustness to noise and the connection between the error bound and the fraction of hyperplanes that separate points of the one-bit CS.

Now, consider the binary measurements, which correspond to the one-bit CS, of the hybrid CS.
Note that (\ref{eq:binary_measurements}) can be rewritten in the form as
\begin{equation}
    [\vy_o]_i\cdot\modd{\va_{o,i},\tilde{\vx}}\geq 0,i=1,2,...,m_o.
    \label{eq:Binary_ineq}
\end{equation}
Assume that the original signal is $\vx\in\cK_{n,s}$. Suppose that there exists an estimate $\hat{\vx}\in\cK_{n,s}$ to $\vx$, then the error bound between $\vx$ and $\hat{\vx}$ can be analyzed, as shown in Theorem 4.

\textit{Theorem 4 (Random uniform tessellations for sparse signals): Consider a matrix $\mA_o\in\bR^{m_o\times n}$, which has independent standard normal entries divided by $\sqrt{m_o}$. Let $0<\delta<1$.
Suppose that $n \geq 2 m_o$, $s \geq C m_o$, and
\begin{equation}
    m_o\geq C'\delta^{-4}s\mbox{log}(n/s).
\end{equation}
Then, with probability at least $1-3\mbox{exp}(-c'\delta^4 m_o)$ over the choice of $\mA_o$,
for any $\vx\in\cK_{n,s}$ and $\ve_o\in\bR^{m_o}$ with $\|\ve_o\|_2\leq c\delta^3 \|\vx\|_2$, the estimate $\hat{\vx}\in\cK_{n,s}$ satisfies
\begin{equation}
        \|\frac{\vx}{\|\vx\|_2}-\frac{\hat{\vx}}{\|\hat{\vx}\|_2}\|_2\leq \delta+d_{\mA_o}(\tilde{\vx},\hat{\vx}).
        \label{eq:x_tilde}
    \end{equation}
The above $C$, $C'$, $c$, and $c'$ are positive constants.
}

\begin{proof}
See Appendix \ref{pf:Thm4}.
\end{proof}

From Theorem 4, several properties of the one-bit CS can be observed. Note that the error bound (\ref{eq:x_tilde}) can be lowered by decreasing $d_{\mA_o}(\tilde{\vx},\hat{\vx})$, \emph{i.e.}, the estimate $\hat{\vx}$, in replace of $\tilde{\vx}$, satisfies more binary inequalities in (\ref{eq:Binary_ineq}).
In addition, conditioned on the fixed lower bound of the success probability in Theorem 4, the value $\delta$ can be a smaller one if $m_o$ increases, resulting in a lower error bound.




\section{Proposed Algorithms for Hybrid CS}
\label{sec:method}
In the previous section, it is shown that, in the one-bit CS, the error bound between a sparse signal and its sparse estimate can be lowered by decreasing the fraction of hyperplanes generated from the measurement matrix that separate the contaminated signal and the estimate.
In addition, the one-bit CS is proved to be optimal for direction estimation of signals under noise with a fixed bit budget \cite{slawski18}.

Note that the traditional CS preserves the energy information of signals in its linear measurements. Therefore, the traditional CS can provide residue information and estimated signals.

In this section, we propose two greedy algorithms for the hybrid CS based on the above discussions.
Note that in the hybrid CS, one would obtain the noisy linear measurements $\vy_r$ and the noisy binary measurements $\vy_o$, which are defined in (\ref{eq:linear_measurements}) and (\ref{eq:binary_measurements}), respectively.
The proposed greedy algorithms integrate the advantages of both types of CS, with the one-bit CS responsible for support detection and the traditional CS offering updated residues and signal estimates.
\subsection{Binary Inequality Checking with Residue Update}

In this subsection, we introduce the first algorithm (see Algorithm 1), which detects the support of the original signal $\vx$ in a greedy manner, and provide a theorem regarding the probability of successful detection.

Algorithm 1 requires the following inputs: the measurement matrix $\mA_r$ and the linear measurements $\vy_r$ of the traditional CS, the measurement matrix $\mA_o$ and the binary measurements $\vy_o$ of the one-bit CS, and the sparsity level $s$ of the original signal $\vx$. During the initialization stage, the support is initialized as an empty set, and the residue is set as the linear measurements $\vy_r$ of the traditional CS.

In each iteration, Algorithm 1 will pick up a specific index from $\{1,2,...,n\}$ in a greedy manner and add it into the estimated support set. The goal of this algorithm is to detect the support $\Omega$ of the original signal $\vx$ after $s$ iterations.

During the $j$th iteration, $j=1,2,...,s$, Algorithm 1 will execute the following three main steps:\\
(Step 1) Candidate selection: The detected support set obtained in the $(j-1)$th iteration is denoted as $\Omega_{j-1}$. For each of the indexes in $\Omega_{j-1}^c$, compute the magnitude of the inner product between the corresponding column of the matrix $\mA_r$ and the residue $\vr_{j-1}$ obtained in the $(j-1)$th iteration individually. Then, sort these values in the decreasing order into a list, and select the column indexes corresponding to the $\floor{\frac{s-j+1}{s}n}$ largest values as candidates $\cT_j$ in this iteration. Note that we contract the number of candidates $\floor{\frac{s-j+1}{s}n}$ as the iterations proceed (with $j$ increasing) in accordance with the decreasing number of undetected support positions. For notational brevity, we define $\kappa_j:=\floor{\frac{s-j+1}{s}n}$, and we will use $\kappa_j$ instead of $\floor{\frac{s-j+1}{s}n}$ hereafter.\\
(Step 2) Support detection: Now we have the $\kappa_j$ candidates in $\cT_j$, where we would like to choose an index $\hat{p}$, whose corresponding estimate $\tau_n(\mA_{r_{\{\Omega_{j-1} \cup \hat{p}\}}}^{\dagger}\vy_r,\{\Omega_{j-1} \cup \hat{p}\})$, in replace of $\tilde{\vx}$, satisfies the most binary inequalities in (\ref{eq:Binary_ineq}), as the index detected in this iteration. Then, we add $\hat{p}$ into the detected support set and obtain $\Omega_j$.\\
(Step 3) Residue update: With the detected support set $\Omega_j$, we derive the residue $\vr_j$ for this iteration.

The following theorem provides an analysis of Algorithm 1 on detecting $\Omega$ after $s$ iterations.

\textit{Theorem 5 (Successful detection with Algorithm 1): Let $\mA_r\in\bR^{m_r\times n}$ and $\mA_o\in\bR^{m_o\times n}$ have independent standard normal entries, which are divided by $\sqrt{m_r}$ and $\sqrt{m_o}$, respectively. Consider any $\vx\in\Sigma_{n,s}$ whose support $\Omega=\{i_1,i_2,...,i_s\}$ satisfies $|[\vx]_{i_1}|\geq|[\vx]_{i_2}|\geq...\geq|[\vx]_{i_s}|$. Assume that $\mA_r$ satisfies $\delta_n\in(0,0.5]$ and $\|\ve_r\|_2\leq\frac{\|\vx\|_2-\sqrt{2}\|[\vx]_{\Omega\setminus i_1}\|_2}{2+\sqrt{2}}$ with $\frac{\|\vx\|_2}{\|[\vx]_{\Omega\setminus i_1}\|_2}\geq\sqrt{2}$. Let $c_j>0$ and $n_j\in\bN$ be the reference value of candidate selection and the threshold number of support detection in the $j$th iteration.
Then, Algorithm 1 detects $\Omega$ after $s$ iterations with probability at least}
\begin{align}
    \prod_{j=2}^s & (1-\sqrt{\frac{2}{\pi}}c_j) [1-I_{e^{-\frac{{c_j}^2}{2}}}(\kappa_j,n-j+1-\kappa_j)]\nonumber\\
    \times\prod_{j=1}^s & \{ 1-I_{\theta_{j,1}}(m_o-n_j+1,n_j)\nonumber\\
    & -(\kappa_j-1)[1-I_{\theta_{j,2}}(m_o-n_j+1,n_j)]\}
\end{align}
\textit{where} $\theta_{j,1}=\frac{\mbox{sin}^{-1}(1-\frac{\sqrt{2}(\|[\vx]_{\Omega\setminus i_1}\|_2-\|[\vx]_{\Omega\setminus\{i_1,i_2,...,i_j\}}\|_2)}{\|\vx\|_2})}{\pi}$ \textit{and} $\theta_{j,2}=\frac{\mbox{cos}^{-1}(\frac{\|[\vx]_{\{i_1,i_2,...,i_{j-1},i_{j+1}\}}\|_2}{\|\vx\|_2})}{\pi}$.

\begin{proof}
See Appendix \ref{pf:Thm5}.
\end{proof}

\vspace{-0.3cm}
\begin{algorithm}
\SetAlgoLined
\textbf{Input:} $\mA_r$, $\mA_o$, $\vy_r$, $\vy_o$, $s$

\textbf{Initialization:} $\Omega_0=\varnothing$, $\vr_0=\vy_r$

\textbf{for} $j=1:s$

1) $\cT_j=\mbox{supp}([\mA_{r_{[\Omega_{j-1}:0]}}^*\vr_{j-1}]^{\floor{\frac{s-j+1}{s}n}})$


2) $\Omega_j=\Omega_{j-1}\cup\underset{\hat{p}\in\cT_j}{\mbox{argmax}}|\{i \mid [\vy_o]_i\cdot[\hat{\mA}_{j,\hat{p}}\vy_r]_i\geq 0,i=1,2,...,m_o\}|$, where $\hat{\mA}_{j,\hat{p}}=\mA_{o_{\{\Omega_{j-1}\cup \hat{p}\}}}\mA_{r_{\{\Omega_{j-1}\cup \hat{p}\}}}^{\dagger}\in\bR^{m_o\times m_r}$

3) $\vr_j=\vy_r-\mA_{r_{\Omega_j}}\mA_{r_{\Omega_j}}^{\dagger}\vy_r$

\textbf{end for}

$\hat{\vx}=\tau_n(\mA_{r_{\Omega_j}}^{\dagger}\vy_r,\Omega_j)$

\textbf{Return:} $\hat{\vx}$

\caption{Binary Inequality Checking with Residue Update}
\end{algorithm}
\vspace{-0.7cm}
\subsection{Support Modification via Binary Inequality Checking}
In this subsection, we will present the second algorithm, as described in Algorithm 2, which refines an initial guess of the support of the original signal $\vx$ in a greedy manner until convergence, and provide a theorem regarding the probability of successful modification.

Algorithm 2 requires the measurement matrix $\mA_r$ and the linear measurements $\vy_r$ of the traditional CS, the measurement matrix $\mA_o$ and the binary measurements $\vy_o$ of the one-bit CS, the sparsity level $s$ of the original signal $\vx$, and an initial guess of the support of $\vx$, $\Omega_0$, with $|\Omega_0|=s$ as inputs. During the initialization stage, we only set the iteration index $j$ as 1, and there is no need to initialize any other parameter.

In each iteration, we will first select a new index and add it into the estimated support set, and then eliminate an index from the estimated support set. As a result, the cardinality of the estimated support set will always be equal to $s$ after arbitrary times of iterations.

During the $j$th iteration, $j\in\bN$, Algorithm 2 executes the following two main steps:\\
(Step 1) Support augmentation: With the estimated support set $\Omega_{j-1}$ obtained in the $(j-1)$th iteration, we would like to choose an index $\hat{p}\in\Omega_{j-1}^c$, whose corresponding estimate $\tau_n(\mA_{r_{\{\Omega_{j-1} \cup \hat{p}\}}}^{\dagger}\vy_r,\{\Omega_{j-1} \cup \hat{p}\})$, in replace of $\tilde{\vx}$, satisfies the most binary inequalities in (\ref{eq:Binary_ineq}), from the $|\Omega_{j-1}^c|=n-s$ candidates, as the index appended to the estimated support set. Consequently, we would obtain the augmented estimated support set $\cT_j$.\\
(Step 2) Support pruning: Now we have the augmented estimated support set $\cT_j$ with $|\cT_j|=s+1$. Then, we would like to obtain a pruned estimated support set $\Omega_j\subset\cT_j$ with $|\Omega_j|=s$, from the $\binom{s+1}{s}=s+1$ candidates, such that $\tau_n(\mA_{r_{\Omega_j}}^{\dagger}\vy_r,\Omega_j)$, in replace of $\tilde{\vx}$, satisfies the most binary inequalities in (\ref{eq:Binary_ineq}).

Note that the stopping criterion of Algorithm 2 is that $\Omega_j=\Omega_{j-1}$, which results in a convergence.
In the following, Theorem 6 provides an analysis of Algorithm 2 on refining the initial guess $\Omega_0$ of the support of $\vx$ into $\Omega$.

\textit{Theorem 6 (Successful Modification with Algorithm 2): Let $\mA_r\in\bR^{m_r\times n}$ and $\mA_o\in\bR^{m_o\times n}$ have independent standard normal entries, which are divided by $\sqrt{m_r}$ and $\sqrt{m_o}$, respectively. Consider any $\vx\in\Sigma_{n,s}$ whose support $\Omega=\{i_1,i_2,...,i_s\}$.
Suppose that the initial guess $\Omega_0$ fails to detect $s'$ support indexes $\{i_1,i_2,...,i_{s'}\}$ with $|[\vx]_{i_1}|\geq|[\vx]_{i_2}|\geq...\geq|[\vx]_{i_{s'-1}}|\geq|[\vx]_{i_{s'}}|$, 
i.e., the set of initial detected support indexes is $\tilde{\Omega}_0=\{i_{s'+1},i_{s'+2},...,i_{s'}\}$.
Assume that $\mA_r$ satisfies $\delta_n\in(0,0.5]$ and $\|\ve_r\|_2\leq\frac{\|\vx\|_2-\sqrt{2}\|[\vx]_{\Omega\setminus\{\tilde{\Omega}_0\cup i_1\}}\|_2}{2+\sqrt{2}}$ with $\frac{\|\vx\|_2}{\|[\vx]_{\Omega\setminus\{\tilde{\Omega}_0\cup i_1\}}\|_2}\geq\sqrt{2}$. Let $\hat{n}_j,\tilde{n}_j\in\bN$ be the threshold number of support augmentation and support pruning, respectively, in the $j$th iteration.
Then, Algorithm 2 modifies $\Omega_0$ into $\Omega$ after $s'$ iterations with probability at least}
\vspace{-0.32cm}
\begin{align}
    \prod_{j=1}^s & \{1-I_{\hat{\theta}_{j,1}}(m_o-\hat{n}_j+1,\hat{n}_j)-(n-s-1)\times\nonumber\\
    & [1-I_{\hat{\theta}_{j,2}}(m_o-\hat{n}_j+1,\hat{n}_j)]\}\times\{1-(s-s'+j)\times\nonumber\\ 
    & I_{\hat{\theta}_{j,1}}(m_o-\tilde{n}_j+1,\tilde{n}_j)-(s'-j+1)\times\nonumber\\
    & [1-I_{\tilde{\theta}_{j,2}}(m_o-\tilde{n}_j+1,\tilde{n}_j)]\}.
\end{align}
\vspace{-0.1cm}
\textit{where} $\hat{\theta}_{j,1}=\frac{\mbox{sin}^{-1}(1-\frac{\sqrt{2}(\|[\vx]_{\Omega\setminus\{\tilde{\Omega}_0\cup i_1\}}\|_2-\|[\vx]_{\Omega\setminus\{\tilde{\Omega}_0,i_1,i_2,...,i_{j-1}\}}\|_2)}{\|\vx\|_2})}{\pi}$, $\hat{\theta}_{j,2}=\frac{\mbox{cos}^{-1}(\frac{\|[\vx]_{\{\tilde{\Omega}_0,i_1,i_2,...,i_{j-1},i_{j+1}\}}\|_2}{\|\vx\|_2})}{\pi}$\textit{, and} $\tilde{\theta}_{j,2}=\frac{\mbox{cos}^{-1}(\frac{\|[\vx]_{\{\tilde{\Omega}_0,i_1,i_2,...,i_j\}\setminus i_j^*}\|_2}{\|\vx\|_2})}{\pi}$ with $i_j^*=\underset{i\in\{\tilde{\Omega}_0,i_1,i_2,...,i_j\}}{\mbox{argmin}}|[\vx]_i|$.
\vspace{-0.1cm}
\begin{proof}
See Appendix \ref{pf:Thm6}.
\end{proof}

\vspace{-0.48cm}
\begin{algorithm}
\SetAlgoLined
\textbf{Input:} $\mA_r$, $\mA_o$, $\vy_r$, $\vy_o$, $s$, $\Omega_0$

\textbf{Initialization:} $j=1$

\textbf{while not converge}

1) $\cT_j=\Omega_{j-1}\cup\underset{\hat{p}\in\Omega_{j-1}^c}{\mbox{argmax}}|\{i \mid [\vy_o]_i\cdot[\hat{\mA}_{j,\hat{p}}\vy_r]_i\geq 0,i=1,2,...,m_o\}|$, where $\hat{\mA}_{j,\hat{p}}=\mA_{o_{\{\Omega_{j-1}\cup \hat{p}\}}}\mA_{r_{\{\Omega_{j-1}\cup \hat{p}\}}}^{\dagger}\in\bR^{m_o\times m_r}$

2) $\Omega_j=\underset{\cT\subset\cT_j, |\cT|=s}{\mbox{argmax}}|\{i \mid [\vy_o]_i\cdot[\hat{\mA}_{\cT}\vy_r)]_i\geq 0,i=1,2,...,m_o\}|$, where $\hat{\mA}_{\cT}=\mA_{o_{\cT}}\mA_{r_{\cT}}^{\dagger}\in\bR^{m_o\times m_r}$

3) $j\leftarrow j+1$

\textbf{end while}

$\hat{\vx}=\tau_n(\mA_{r_{\Omega_j}}^{\dagger}\vy_r,\Omega_j)$

\textbf{Return:} $\hat{\vx}$

\caption{Support Modification via Binary Inequality Checking}
\end{algorithm}

\section{Simulation}
\label{sec:simulation}
In this section, we present the numerical results in order to compare the recovery performances of the proposed greedy algorithms of hybrid CS with those of the classic greedy algorithms of traditional CS for two experiments with each algorithm sharing the same bit budget.
Note that the greedy algorithms of one-bit CS are not considered for the simulations since they require the prior knowledge of the signals' energy information, which is hardly known in real-world applications.
The greedy algorithms of traditional CS that are included in the experiments are the orthogonal matching pursuit (OMP) \cite{tropp07}, subspace pursuit (SP) \cite{dai09}, and CoSaMP \cite{NEEDELL09}.

\begin{figure}%
\centering
\subfigure[$s=4$]{%
\label{fig:Exp1_s4}%
\includegraphics[width=8.5cm]{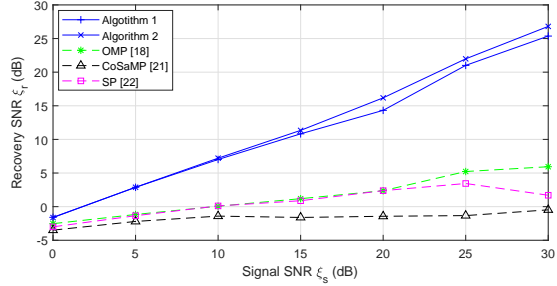}}%
\qquad
\subfigure[$s=8$]{%
\label{fig:Exp1_s8}%
\includegraphics[width=8.5cm]{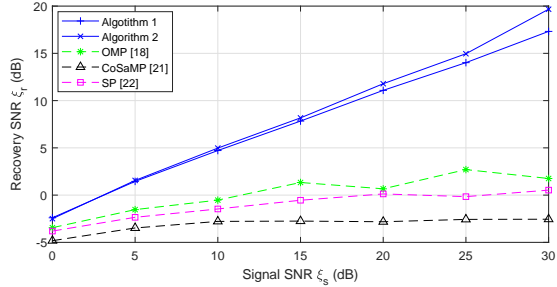}}%
\qquad
\subfigure[$s=16$]{%
\label{fig:Exp1_s16}%
\includegraphics[width=8.5cm]{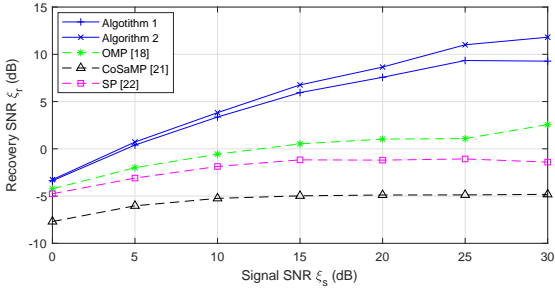}}%
\qquad
\subfigure[$s=32$]{%
\label{fig:Exp1_s32}%
\includegraphics[width=8.5cm]{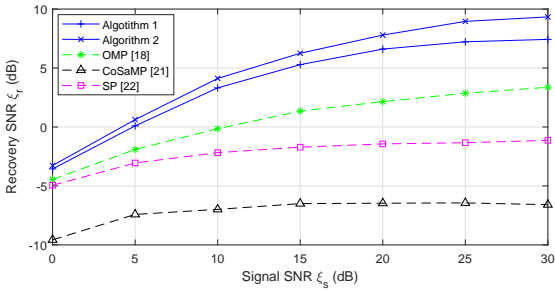}}%
\caption{Recovery performances of Experiment 1}
\label{fig:Exp1}
\end{figure}

\begin{figure}%
\centering
\subfigure[$\xi_s=0$ dB]{%
\label{fig:Exp2_0dB}%
\includegraphics[width=8.5cm]{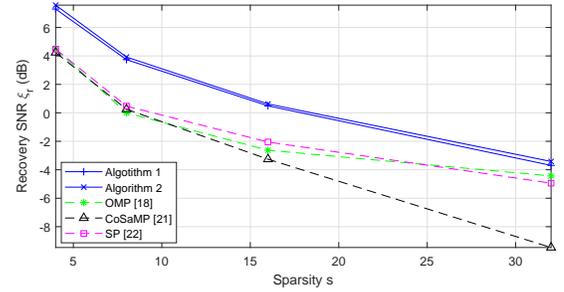}}%
\qquad
\subfigure[$\xi_s=10$ dB]{%
\label{fig:Exp2_10dB}%
\includegraphics[width=8.5cm]{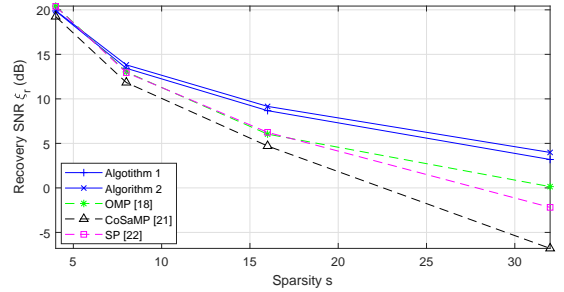}}%
\caption{Recovery performances of Experiment 2}
\label{fig:Exp2}
\end{figure}
\vspace{-0.2cm}
\subsection{Parameter Settings}
Each linear measurement in traditional CS is stored as a float number which consumes 32 bits, while each binary measurement in one-bit CS consumes only 1 bit.
For each test of each experiment, we adopt the Monte Carlo simulation by randomly generating $n_v=500$ $s$-sparse vectors of dimension $n=256$, with each vector containing standard normal entries in its support, as the signal $\vx$ to be recovered.
With each signal generated, we construct corresponding measurement matrices $\mA_r\in\bR^{m_r\times n}$ and $\mA_o\in\bR^{m_o\times n}$ for the proposed greedy algorithms of hybrid CS, and $\mA\in\bR^{m\times n}$ for the greedy algorithms of traditional CS. Note that these measurement matrices contain the standard normal entries divided by the square root of the number of measurements.
Also, Algorithm 2 takes the estimated support set obtained from Algorithm 1 as the initial guess.
There are two types of signal-to-noise ratio (SNR) used in the simulations, signal SNR $\xi_s:=10\mbox{log}\frac{\|\vx\|_2^2}{\|\vu\|_2^2}$ (dB) and recovery SNR $\xi_r:=10\mbox{log}\frac{1}{n_v}\sum_{i=1}^{n_v}\frac{\|\vx_i\|_2^2}{\|\vx_i-\hat{\vx}_i\|_2^2}$ (dB), where $\vx_i$ and $\hat{\vx}_i$ are the $i$th signal and estimate in a single test.
Note that the recovery performances are evaluated in terms of $\xi_r$. 

The first experiment evaluates the recovery performances of all greedy algorithms under a relatively small budget of $64s$ bits in the noisy case with the tests executed across $s=\{4,8,16,32\}$ and $\xi_s=\{0,5,10,15,20,25,30\}$.
In the first experiment, $m_r=\ceil{1.5s}$ and $m_o=32\times\floor{0.5s}$ for the greedy algorithms of hybrid CS, and $m=2s$ for those of traditional CS.

The second experiment evaluates the recovery performances of all greedy algorithms under a fixed budget of $64\times32=2048$ bits in the noisy case with the tests executed across $s=\{4,8,16,32\}$ and $\xi_s=\{0,10\}$.
In the second experiment, $m_r=48$ and $m_o=16\times32=512$ for the greedy algorithms of hybrid CS, and $m=64$ for those of traditional CS.

\subsection{Simulation Results}
For the first experiment, the simulation results are shown in Fig. \ref{fig:Exp1}, where we compare the recovery performances of the proposed greedy algorithms of hybrid CS with those of the greedy algorithms of traditional CS in the noisy case under a relatively small bit budget.
It can be observed that the proposed greedy algorithms outperform the greedy algorithms of traditional CS, which shows the robustness to noise and the feasibility under a small bit budget of the proposed algorithms.
This is because the proposed algorithms strike a balance between the pros and cons of both the traditional CS and one-bit CS.
Particularly, the traditional CS can be regarded as a special case of the hybrid CS.
Note that the recovery performances of Algorithm 2 is better than those of Algorithm 1, which demonstrates the efficacy of Algorithm 2 on modifying the initial guess of the original signal's support into a more correct one.
In addition, it can be found that the recovery performances of all greedy algorithms degrade when $s$ increases or when the signal SNR decreases.

For the second experiment where the bit budget is fixed, the simulation results are shown in Fig. \ref{fig:Exp2} and exhibit a similar trend to the first experiment.

\vspace{-0.35cm}
\section{Conclusions}
\label{sec:conclusion}
\vspace{-0.15cm}
In this study, we focus on the hybrid compressed sensing (CS), which combines both the traditional CS and one-bit CS.
First, we propose a fundamental theorem on the random uniform tessellations for sparse signals in order to investigate the properties of one-bit CS.
Then, we propose two greedy algorithms for the hybrid CS, with the one-bit CS detecting the support indexes and the traditional CS providing the updated residues and signal estimates.
For each of the two algorithms, we offer the corresponding theorem to theoretically analyze its capabilities.
Numerical results have demonstrated the efficacy of the proposed two algorithms, compared with classic greedy algorithms of traditional CS.
The future directions of the hybrid CS include the optimal split of bit budget for linear measurements and binary measurements, and Bayesian algorithms with statistical methods.



\section*{Acknowledgment}
\vspace{-0.1cm}
This work was supported by Ministry of Science and Technology, Taiwan, under Grants MOST 107-2221-E-001-015-MY2 and MOST 108-2634-F-007-010.
\vspace{-0.4cm}
\begin{appendices}

\section{Proof of Theorem 4}
\label{pf:Thm4}
First, note that $\|\ve_o\|_2\leq c \delta^3\|\vx\|_2$.
According to Corollary 1, with failure probability at most $\mbox{exp}(-c m_o)$, there exists a vector $\vu$ satisfying
    \begin{equation}
        \ve_o=\mA_o\vu \quad \mbox{with}
\left\{
\begin{array}{l}
\|\vu\|_2 \leq \frac{\delta}{\delta+3}\|\vx\|_2\\
\|\vu\|_1 \leq \frac{\sqrt{\mbox{log}2\cdot C}}{2} \sqrt{m_o/\mbox{log}(n/m_o)}\|\vx\|_2. 
\label{eqs:l1l2}
\end{array}
\right .
    \end{equation}
Note that $n/m_o\geq 2$, and thus $\mbox{log}(n/m_o)\geq \mbox{log}2$. Therefore, 
\begin{align}
    \|\vu\|_1 & \leq \frac{\sqrt{\mbox{log}2\cdot C}}{2} \sqrt{m_o/\mbox{log}(n/m_o)}\|\vx\|_2\nonumber\\
    & \leq\frac{\sqrt{C m_o}}{2}\|\vx\|_2\leq\frac{\sqrt{s}}{2}\|\vx\|_2,
\end{align}
where the last inequality holds since $s\geq Cm_o$.

With the triangular inequality, the upper bound and the lower bound of the $\ell_2$-norm of the contaminated signal $\tilde{\vx}$ can be derived as
\begin{equation}
    \|\tilde{\vx}\|_2\leq\|\vx\|_2+\|\vu\|_2\leq(1+\frac{\delta}{\delta+3})\|\vx\|_2\leq\frac{5}{4}\|\vx\|_2,
    \label{eq:tilde2a}
\end{equation}
\begin{equation}
        \|\tilde{\vx}\|_2\geq \|\vx\|_2-\|\vu\|_2\geq (1-\frac{\delta}{\delta+3})\|\vx\|_2\geq\frac{3}{4}\|\vx\|_2.
        \label{eq:tilde2}
\end{equation}
In a similar manner, the upper bound of the $\ell_1$-norm of $\tilde{\vx}$ can be derived as
\begin{equation}
         \|\tilde{\vx}\|_1\leq \|\vx\|_1+\|\vu\|_1 \leq (\sqrt{s}+\frac{\sqrt{s}}{2})\|\vx\|_2=\frac{3}{2}\sqrt{s}\|\vx\|_2.
         \label{eq:tilde1}
\end{equation}
Combining (\ref{eq:tilde2}) and (\ref{eq:tilde1}), the upper bound of the $\ell_1$-norm of the normalized contaminated signal $\frac{\tilde{\vx}}{\|\tilde{\vx}\|_2}$ can be obtained as
\begin{equation}
    \|\frac{\tilde{\vx}}{\|\tilde{\vx}\|_2}\|_1\leq\frac{\frac{3}{2}\sqrt{s}\|\vx\|_2}{\frac{3}{4}\|\vx\|_2}=2\sqrt{s}.
\end{equation}
Therefore, the normalized contaminated signal $\frac{\tilde{\vx}}{\|\tilde{\vx}\|_2}\in 2\sqrt{s}\cB^n_1\cap\cS^{n-1}$.
Note that $\frac{\hat{\vx}}{\|\hat{\vx}\|_2}\in\sqrt{s}\cB_1^n\cap\cS^{n-1}$, and therefore $\frac{\hat{\vx}}{\|\hat{\vx}\|_2}\in 2\sqrt{s}\cB_1^n\cap\cS^{n-1}$.

According to (\ref{eq:GMW_import}) and Theorem 3, with probability at least $1-2\mbox{exp}(-c\delta^4 m_o)$, the following satisfies:
    \begin{align}
        & \|\frac{\tilde{\vx}}{\|\tilde{\vx}\|_2}-\frac{\hat{\vx}}{\|\hat{\vx}\|_2}\|_2\leq\frac{\delta}{3}+d_A(\frac{\tilde{\vx}}{\|\tilde{\vx}\|_2},\frac{\hat{\vx}}{\|\hat{\vx}\|_2})=\frac{\delta}{3}+d_A(\tilde{\vx},\hat{\vx}),\nonumber\\
        & \frac{\tilde{\vx}}{\|\tilde{\vx}\|_2},\frac{\hat{\vx}}{\|\hat{\vx}\|_2}\in 2\sqrt{s}\cB_1^n\cap\cS^{n-1},
        \label{eq:ineq1}
    \end{align}
    given $m_o\geq C'\delta^{-4}s\mbox{log}(n/s)$.
    
We target the inequality
\begin{align}
    \|\frac{\vx}{\|\tilde{\vx}\|_2}-\frac{\tilde{\vx}}{\|\tilde{\vx}\|_2}\|_2=\|\frac{\vu}{\|\tilde{\vx}\|_2}\|_2  & \leq \|\frac{\vu}{(1-\frac{\delta}{\delta+3})\|\vx\|_2}\|_2\nonumber\\
    & \leq\frac{\frac{\delta}{\delta+3}\|\vx\|_2}{\frac{3}{\delta+3}\|\vx\|_2}=\frac{\delta}{3},
    \label{eq:ineq2}
\end{align}
where the first inequality holds due to the inequality (\ref{eq:tilde2}), and the last inequality holds due to the upper bound of $\|\vu\|_2$ obtained in (\ref{eqs:l1l2}).
In addition, it can be derived that
\begin{align}
    \|\frac{\vx}{\|\vx\|_2}-\frac{\vx}{\|\tilde{\vx}\|_2}\|_2 & \leq \mbox{max}(|1-\frac{1}{1+\frac{\delta}{\delta+3}}|,|1-\frac{1}{1-\frac{\delta}{\delta+3}}|)\nonumber\\
    & =\mbox{max}(\frac{\delta}{2\delta+3},\frac{\delta}{3})=\frac{\delta}{3},
    \label{eq:ineq3}
\end{align}
where the first inequality holds due to the upper bound and lower bound of $\|\tilde{\vx}\|_2$ as stated in (\ref{eq:tilde2a}) and (\ref{eq:tilde2}), and the last equality holds due to the fact that $\delta$ is chosen from the interval $(0,1]$.

Finally, with the triangular inequality, we obtain that
\begin{align}
    \|\frac{\vx}{\|\vx\|_2}-\frac{\hat{\vx}}{\|\hat{\vx}\|_2}\|_2
    & \leq\|\frac{\vx}{\|\vx\|_2}-\frac{\vx}{\|\tilde{\vx}\|_2}\|_2+\|\frac{\vx}{\|\tilde{\vx}\|_2}-\frac{\tilde{\vx}}{\|\tilde{\vx}\|_2}\|_2\nonumber\\
    & +\|\frac{\tilde{\vx}}{\|\tilde{\vx}\|_2}-\frac{\hat{\vx}}{\|\hat{\vx}\|_2}\|_2\leq \delta+d_A(\tilde{\vx},\hat{\vx}),
\end{align}
where the first inequality holds due to the triangular inequality, and the last inequality holds due to the inequalities (\ref{eq:ineq1}), (\ref{eq:ineq2}), and (\ref{eq:ineq3}).

\section{Proof of Theorem 5}
\label{pf:Thm5}
Without loss of generality, assume that the first $s$ entries of $\vx$ are nonzero with $|[\vx]_1|\geq |[\vx]_2|\geq ... \geq |[\vx]_s|$, while the remaining $n-s$ entries are equal to zero. Therefore, we denote the support set and the non-support set of $\vx$ as $\Omega=\{1,2,...,s\}$ and $\cN\cS=\{s+1,s+2,...,n\}$, respectively.

Now, we define two events, $E_{\footnotesize{\mbox{succ}}}$: the support set $\Omega$ is detected after the $s$ iterations and $E_{\varnothing\rightarrow a'}$: the order of detected indexes after the first $a'$ iterations is $1\rightarrow 2\rightarrow ... \rightarrow a'$.
Consider a specific event $E_{\varnothing \rightarrow s}$ where the order of detected indexes after the $s$ iterations is: $1\rightarrow 2\rightarrow ... \rightarrow s$. Note that when the event $E_{\varnothing \rightarrow s}$ happens, the event $E_{\footnotesize{\mbox{succ}}}$ also happens. Therefore, it can be inferred that
\begin{equation}
    \mathbb{P}\{E_{\footnotesize{\mbox{succ}}}\}\geq\mathbb{P}\{E_{\varnothing \rightarrow s}\}.
    \label{eq:succ}
\end{equation}
With the multiplication law of probability, $\mathbb{P}\{E_{\varnothing \rightarrow s}\}$ can be decomposed as
\begin{align}
    \mathbb{P}\{E_{\varnothing \rightarrow s}\} & =\mathbb{P}\{E_{\varnothing \rightarrow 1}\}\times\mathbb{P}\{E_{1 \rightarrow 2}\mid E_{\varnothing \rightarrow 1}\}\times...\nonumber\\
    & \times\mathbb{P}\{E_{j-1 \rightarrow j}\mid E_{\varnothing \rightarrow j-1}\}\times...\nonumber\\
    & \times\mathbb{P}\{E_{s-1 \rightarrow s}\mid E_{\varnothing \rightarrow s-1}\},
    \label{eq:cond}
\end{align}
where the event $E_{a'-1\rightarrow a'}$ is defined as: the detected index in the $(a'-1)$th iteration is $a'-1$ and the detected index in the $a'$th iteration is $a'$.

Note that the first term on the RHS of (\ref{eq:cond}) corresponds to the detection of the index 1 in the first iteration, while the $a'$th term on the RHS of (\ref{eq:cond}) corresponds to the detection of the index $a'$ in the $a'$th iteration, given that the order of detected indexes in the previous $a'-1$ iterations is: $1\rightarrow 2\rightarrow ...\rightarrow a'-1$, $a'=2,3,...,s$.

Define two events, $E_{\footnotesize{\mbox{cand}},a'}$: the index $a'$ is selected as one of the $\kappa_{a'}$ candidates in $\cT_{a'}$ in the $a'$th iteration and $E_{\footnotesize{\mbox{supp}},a'}$: the index $a'$ is detected in the $a'$th iteration. Then, with the multiplication law of probability, the first term and the $a'$th term, $a'=2,3,...,s$, on the RHS of (\ref{eq:cond}) can be decomposed as
\begin{align}
    \mathbb{P}\{E_{\varnothing \rightarrow 1}\}= & \mathbb{P}\{E_{\footnotesize{\mbox{cand}},1}\cap E_{\footnotesize{\mbox{supp}},1}\}\nonumber\\
    & \mathbb{P}\{E_{\footnotesize{\mbox{cand}},1}\}\times\mathbb{P}\{E_{\footnotesize{\mbox{supp}},1}\mid E_{\footnotesize{\mbox{cand}},1}\},
    \label{eq:j1}
\end{align}
\begin{align}
     \mathbb{P}\{E_{a'-1 \rightarrow a'}\mid E_{\varnothing \rightarrow a'-1}\} & =\mathbb{P}\{E_{\footnotesize{\mbox{cand}},a'}\cap E_{\footnotesize{\mbox{supp}},a'}\mid E_{\varnothing \rightarrow a'-1}\}\nonumber\\
     & =\mathbb{P}\{E_{\footnotesize{\mbox{cand}},a'}\mid E_{\varnothing \rightarrow a'-1}\}\nonumber\\
     & \times\mathbb{P}\{E_{\footnotesize{\mbox{supp}},a'}\mid E_{\footnotesize{\mbox{cand}},a'},E_{\varnothing \rightarrow a'-1}\},
     \label{eq:jn}
\end{align}
respectively.
In the following proof, we will address the probability terms on the RHS of (\ref{eq:j1}) and (\ref{eq:jn}).

First, consider the step of candidate selection. In the $j$th iteration, we would select the $\kappa_j$ candidates from $\Omega_{j-1}^c$ as $\cT_j$, which should contain the index $j$ so that we may detect it in the next step.

When $j=1$, we would select $\kappa_1=n$ candidates from $\Omega_0^c=\{1,2,...,n\}$. It is obvious that the index 1 must be chosen as one of the candidates in $\cT_1$. Therefore, we derive that
\begin{equation}
    \mathbb{P}\{E_{\footnotesize{\mbox{cand}},1}\}=1.
    \label{eq:cand1}
\end{equation}

When $j=2,3,...,s$, the detected support set obtained in the previous $j-1$ iterations is $\Omega_{j-1}=\{1,2,...,j-1\}$, conditioned on the event $E_{\varnothing\rightarrow j-1}$. Note that we would like to have the index $j$ as one of the candidates in $\cT_j$.
Therefore, we still need to select $\kappa_j-1$ candidates, \emph{i.e.}, eliminate $(n-j)-(\kappa_j-1)=n-j+1-\kappa_j$ indexes, from $\{\Omega_{j-1}\cup j\}^c$.

Now, consider the event $E_{j,c_j}$ where $|\modd{\va_{r,j},\frac{\sqrt{m_r}\vr_{j-1}}{\|\vr_{j-1}\|_2}}|>c_j$ and at least $n-j+1-\kappa_j$ indexes from $\{\Omega_{j-1}\cup j\}^c$ satisfy $|\modd{\va_{r,i},\frac{\sqrt{m_r}\vr_{j-1}}{\|\vr_{j-1}\|_2}}|\leq c_j,i\in\{\Omega_{j-1}\cup j\}^c$, where $c_j$ is the reference value for candidate selection in the $j$th iteration and $\va_{r,i'}$ is the $i'$th column of the matrix $\mA_r$. Conditioned on the event $E_{\varnothing\rightarrow j-1}$, when the event $E_{j,c_j}$ happens, the event $E_{\footnotesize{\mbox{cand}},j}$ also happens. Consequently, it can be inferred that
\begin{equation}
    \mathbb{P}\{E_{\footnotesize{\mbox{cand}},j}\mid E_{\varnothing \rightarrow j-1}\}\geq\mathbb{P}\{E_{j,c_j}\mid E_{\varnothing \rightarrow j-1}\}.
    \label{eq:cond_LB1}
\end{equation}

Note that conditioned on the event $E_{\varnothing \rightarrow j-1}$, $\modd{\va_{r,i'},\frac{\sqrt{m_r}\vr_{j-1}}{\|\vr_{j-1}\|_2}}\sim N(0,1),i'=1,2,...,n$. Therefore, we have
\begin{align}
    & \mathbb{P}\{|\modd{\va_{r,i'},\frac{\sqrt{m_r}\vr_{j-1}}{\|\vr_{j-1}\|_2}}|>c\mid E_{\varnothing \rightarrow j-1}\}\nonumber\\
    = & 1-2\mathbb{P}\{\modd{\va_{r,i'},\frac{\sqrt{m_r}\vr_{j-1}}{\|\vr_{j-1}\|_2}}\in(0,c]\mid E_{\varnothing \rightarrow j-1}\}\nonumber\\
    \geq & 1-2\cdot\frac{c}{\sqrt{2\pi}}=1-\sqrt{\frac{2}{\pi}}c, \label{eq:standard_normal1}\\
    & \mathbb{P}\{|\modd{\va_{r,i'},\frac{\sqrt{m_r}\vr_{j-1}}{\|\vr_{j-1}\|_2}}|\leq c\mid E_{\varnothing \rightarrow j-1}\}\nonumber\\ = & 1-2\mathbb{P}\{\modd{\va_{r,i'},\frac{\sqrt{m_r}\vr_{j-1}}{\|\vr_{j-1}\|_2}}>c\mid E_{\varnothing \rightarrow j-1}\}\nonumber\\
    \geq & 1-2\cdot\frac{1}{2}e^{-\frac{c^2}{2}}=1-e^{-\frac{c^2}{2}},
    \label{eq:standard_normal2}
\end{align}
where the inequalities in (\ref{eq:standard_normal1}) and (\ref{eq:standard_normal2}) hold due to the bounds for standard normal distribution.

Denote the number of indexes in $\{\Omega_{j-1}\cup j\}^c$ that satisfy $|\modd{\va_{r,i},\frac{\sqrt{m_r}\vr_{j-1}}{\|\vr_{j-1}\|_2}}|\leq c_j,i\in\{\Omega_{j-1}\cup j\}^c$ conditioned on the event $E_{\varnothing \rightarrow j-1}$ as $B_{\footnotesize{\mbox{Low}},j}\sim B(n-j,p_{\footnotesize{\mbox{Low}},j})$, where $p_{\footnotesize{\mbox{Low}},j}\geq 1-e^{-\frac{{c_j}^2}{2}}$, which follows (\ref{eq:standard_normal2}).

Now, we can compute the probability
\begin{align}
    \mathbb{P}\{E_{j,c_j}\mid E_{\varnothing \rightarrow j-1}\}= & \mathbb{P}\{|\modd{\va_{r,j},\frac{\sqrt{m_r}\vr_{j-1}}{\|\vr_{j-1}\|_2}}|>c\mid E_{\varnothing \rightarrow j-1}\}\nonumber\\
    \times & \mathbb{P}\{B_{\footnotesize{\mbox{Low}},j}\geq n-j+1-\kappa_j\}\nonumber\\
    \geq & (1-\sqrt{\frac{2}{\pi}}c_j)\nonumber\\
    \times & [1-\mathbb{P}\{B_{\footnotesize{\mbox{Low}},j}\leq n-j-\kappa_j\}]\nonumber\\
    \geq & (1-\sqrt{\frac{2}{\pi}}c_j)\nonumber\\
    \times & [1-I_{e^{-\frac{{c_j}^2}{2}}}(\kappa_j,n-j+1-\kappa_j)],
    \label{eq:Ej_cj}
\end{align}
where the first inequality holds due to (\ref{eq:standard_normal1}).

Next, consider the step of support detection.
In the $j$th iteration, we would like to detect the index $j$ from $\cT_j$, conditioned on the event $E_{\footnotesize{\mbox{cand}},j}$.

Now, we define two events, $E_{\Omega_1,n_1}$: when $j=1$, $|\{i\mid[\vy_o]_i\cdot[\hat{\mA}_{1,1}\vy_r]_i\geq 0,i=1,2,...,m_o\}|\geq n_1$ and $|\{i\mid[\vy_o]_i\cdot[\hat{\mA}_{1,p'}\vy_r]_i\geq 0,i=1,2,...,m_o\}|\leq n_1-1,p'\in\{\Omega_0 \cup 1\}^c$, conditioned on the event $E_{\footnotesize{\mbox{cand}},1}$, and $E_{\Omega_{a'},n_{a'}}$: when $j=a',a'=2,3,...,s$, $|\{i\mid[\vy_o]_i\cdot[\hat{\mA}_{a',a'}\vy_r]_i\geq 0,i=1,2,...,m_o\}|\geq n_{a'}$ and $|\{i\mid[\vy_o]_i\cdot[\hat{\mA}_{a',p'}\vy_r]_i\geq 0,i=1,2,...,m_o\}|\leq n_{a'}-1,p'\in\{\Omega_{a'-1}\cup a'\}^c$, conditioned on the events $E_{\footnotesize{\mbox{cand}},a'}$ and $E_{\varnothing \rightarrow a'-1}$, where $n_j,j=1,2,...,s$ is the threshold number of support detection in the $j$th iteration.

Note that when the event $E_{\Omega_1,n_1}$ happens, the event $E_{\footnotesize{\mbox{supp}},1}$ conditioned on the event $E_{\footnotesize{\mbox{cand}},1}$ also happens, and when the event $E_{\Omega_{a'},n_{a'}}$ happens, the event $E_{\footnotesize{\mbox{supp}},a'}$ conditioned on the events $E_{\footnotesize{\mbox{cand}},a'}$ and $E_{\varnothing \rightarrow a'-1}$ also happens. Therefore, it can be inferred that 
\begin{equation}
    \mathbb{P}\{E_{\footnotesize{\mbox{supp}},1}\mid E_{\footnotesize{\mbox{cand}},1}\}\geq\mathbb{P}\{E_{\Omega_1,n_1}\},
    \label{eq:supp1_LB}
\end{equation}
\begin{equation}
    \mathbb{P}\{E_{\footnotesize{\mbox{supp}},a'}\mid E_{\footnotesize{\mbox{cand}},a'},E_{\varnothing \rightarrow a'-1}\}\geq\mathbb{P}\{E_{\Omega_{a'},n_{a'}}\},
    \label{eq:supp_a'_LB}
\end{equation}
where the event $E_{\Omega_j,n_j}$ implies that the detected support set in the $j$th iteration would be $\Omega_j=\{1,2,...,j\}=\{\Omega_{j-1},j\}$ ($\Omega_0=\varnothing$).

In order to analyze the probability of the hyperplanes generated from $\mA_o$ separating $\vx$ and the estimate $\tau_n(\mA^{\dagger}_{r_{\Omega_j}}\vy_r,\Omega_j)$, compute the $\ell_2$-norm distance between both points as
\begin{equation}
     \|\vx-\tau_n(\mA^{\dagger}_{r_{\Omega_j}}\vy_r,\Omega_j)\|_2\leq\|\vx-\tilde{\vx}\|_2+\|\tilde{\vx}-\tau_n(\mA^{\dagger}_{r_{\Omega_j}}\vy_r,\Omega_j)\|_2,
     \label{eq:distance}
\end{equation}
where the inequality holds due to the triangular inequality.
Note that the first term and the second term on the RHS of (\ref{eq:distance}) correspond to the signal-level noise and the distance between $\tilde{\vx}$ and the estimate.

We can compute the square of the second term on the RHS of (\ref{eq:distance}) as
\begin{align}
        \|\tilde{\vx}-\tau_n(\mA^{\dagger}_{r_{\Omega_j}}\vy_r,\Omega_j)\|_2^2= & \|[\tilde{\vx}]_{\Omega_j^c}\|_2^2+\|[\tilde{\vx}]_{\Omega_j}-\mA^{\dagger}_{r_{\Omega_j}}\mA_r\tilde{\vx}\|_2^2\nonumber\\
        = & \|[\tilde{\vx}]_{\Omega_j^c}\|_2^2+ \|[\tilde{\vx}]_{\Omega_j}- \mA^{\dagger}_{r_{\Omega_j}}\times\nonumber\\
        & (\mA_{r_{\Omega_j}}[\tilde{\vx}]_{\Omega_j}+\mA_{r_{\Omega_j^c}}[\tilde{\vx}]_{\Omega_j^c})\|_2^2\nonumber\\
        = & \|[\tilde{\vx}]_{\Omega_j^c}\|_2^2+\|\mA^{\dagger}_{r_{\Omega_j}}\mA_{r_{\Omega_j^c}}[\tilde{\vx}]_{\Omega_j^c}\|_2^2\nonumber\\
            = & \|[\tilde{\vx}]_{\Omega_j^c}\|_2^2+ \|(\mA^*_{r_{\Omega_j}}\mA_{r_{\Omega_j}})^{-1}\times\nonumber\\
            & \mA^*_{r_{\Omega_j}}\mA_{r_{\Omega_j^c}}[\tilde{\vx}]_{\Omega_j^c}\|_2^2\nonumber\\
            \leq & \|[\tilde{\vx}]_{\Omega_j^c}\|_2^2\nonumber\\
            + & \frac{1}{1-\delta_j}\|\mA^*_{r_{\Omega_j}}\mA_{r_{\Omega_j^c}}[\tilde{\vx}]_{\Omega_j^c}\|_2^2\nonumber\\
            \leq & \|[\tilde{\vx}]_{\Omega_j^c}\|_2^2+\frac{\delta_n}{1-\delta_j}\|[\tilde{\vx}]_{\Omega_j^c}\|_2^2\nonumber\\
            = & (1+\frac{\delta_n}{1-\delta_j})\|[\tilde{\vx}]_{\Omega_j^c}\|_2^2,
        \label{eq:UB_supp}
\end{align}
where the first inequality holds due to Lemma 2 and the second inequality holds due to Lemma 3.

Note that the matrix $\mA_r$ satisfies $\delta_n\in(0,0.5]$ and $\delta_j\leq\delta_n$ (which is implied by Lemma 1). 
According to Definition 1, since $\|\ve_r\|_2\leq\frac{\|\vx\|_2-\sqrt{2}\|[\vx]_{\Omega\setminus 1}\|_2}{2+\sqrt{2}}$, it can be inferred that
\begin{equation}
    \|\vu\|_2\leq\frac{\|\vx\|_2-\sqrt{2}\|[\vx]_{\Omega\setminus 1}\|_2}{1+\sqrt{2}}.
    \label{eq:l2_u}
\end{equation}

With (\ref{eq:UB_supp}), (\ref{eq:l2_u}), and the properties of $\mA_r$, (\ref{eq:distance}) becomes
\begin{align}
        \|\vx-\tau_n(\mA^{\dagger}_{r_{\Omega_j}}\vy_r,\Omega_j)\|_2\leq & \|\vu\|_2+\sqrt{1+\frac{\delta_n}{1-\delta_j}}\|[\tilde{\vx}]_{\Omega_j^c}\|_2\nonumber\\
        = & \|\vu\|_2+\sqrt{1+\frac{\delta_n}{1-\delta_j}}\|[\vx+\vu]_{\Omega_j^c}\|_2\nonumber\\
        \leq & \|\vu\|_2+\sqrt{1+\frac{\delta_n}{1-\delta_j}}(\|[\vx]_{\Omega\setminus\Omega_j}\|_2\nonumber\\
        + & \|[\vu]_{\Omega_j^c}\|_2)\nonumber\\
        \leq & \|\vu\|_2+\sqrt{1+\frac{\delta_n}{1-\delta_j}}(\|[\vx]_{\Omega\setminus\Omega_j}\|_2\nonumber\\
        + & \|\vu\|_2)\nonumber\\
        = & (1+\sqrt{1+\frac{\delta_n}{1-\delta_j}})\|\vu\|_2\nonumber\\
        + & \sqrt{1+\frac{\delta_n}{1-\delta_j}}\|[\vx]_{\Omega\setminus\Omega_j}\|_2\nonumber\\
        \leq & (1+\sqrt{2})\|\vu\|_2+\sqrt{2}\|[\vx]_{\Omega\setminus\Omega_j}\|_2\nonumber\\
        \leq & \|\vx\|_2-\sqrt{2}\times\nonumber\\
        & (\|[\vx]_{\Omega\setminus1}\|_2-\|[\vx]_{\Omega\setminus\Omega_j}\|_2),
        \label{eq:UB_noise}
\end{align}
where the second inequality holds due to the triangular inequality.
Note that $\|\vx\|_2-\sqrt{2}(\|[\vx]_{\Omega\setminus1}\|_2-\|[\vx]_{\Omega\setminus\Omega_j}\|_2)\leq\|\vx\|_2$.

Consider the hyperplane with normal vector $\va_{o,i},i=1,2,...,m_o$, which is the $i$th row of the matrix $\mA_o$. 
First, compute the probability of the hyperplane not separating $\vx$ and the estimate obtained from $\Omega_j$.
According to (\ref{eq:UB_noise}) and Lemma 4, we have
\begin{align}
                & \mathbb{P}\{\mbox{sign}(\modd{\va_{o,i},\vx})= \mbox{sign}(\modd{\va_{o,i},\tau_n(\mA_{r_{\Omega_j}}^{\dagger}\vy_r,\Omega_j)})\}\nonumber\\
                = & 1-\mathbb{P}\{\mbox{sign}(\modd{\va_{o,i},\vx})\neq\mbox{sign}(\modd{\va_{o,i},\tau_n(\mA_{r_{\Omega_j}}^{\dagger}\vy_r,\Omega_j)})\}\nonumber\\
                \geq & 1-\frac{\mbox{sin}^{-1}(\frac{\|\vx\|_2-\sqrt{2}(\|[\vx]_{\Omega\setminus 1}\|_2-\|[\vx]_{\Omega\setminus\Omega_j}\|_2)}{\|\vx\|_2})}{\pi}\nonumber\\
                = & 1-\frac{\mbox{sin}^{-1}(1-\frac{\sqrt{2}(\|[\vx]_{\Omega\setminus 1}\|_2-\|[\vx]_{\Omega\setminus\Omega_j}\|_2)}{\|\vx\|_2})}{\pi}.
\end{align}
Next, compute the probability of the hyperplane not separating $\vx$ and the estimate obtained from $\{\Omega_{j-1}\cup p'\},p'\in\cT_j-j$.
According to Lemma 4, we have
\begin{align}
                & \mathbb{P}\{\mbox{sign}(\modd{\va_{o,i},\vx})=\nonumber\\
                &\quad\mbox{sign}(\modd{\va_{o,i},\tau_n(\mA_{r_{\{\Omega_{j-1}\cup p'\}}}^{\dagger}\vy_r,\{\Omega_{j-1}\cup p'\})})\}\nonumber\\
                = & 1-\mathbb{P}\{\mbox{sign}(\modd{\va_{o,i},\vx})\neq\nonumber\\
                &\qquad\quad\mbox{sign}(\modd{\va_{o,i},\tau_n(\mA_{r_{\{\Omega_{j-1}\cup p'\}}}^{\dagger}\vy_r,\{\Omega_{j-1}\cup p'\})})\}\nonumber\\
                \leq & 1-\frac{\mbox{cos}^{-1}(\frac{\|[\vx]_{\{\Omega_{j-1}\cup j+1\}}\|_2}{\|\vx\|_2})}{\pi}.
\end{align}

Denote $|\{i\mid[\vy_o]_i\cdot[\hat{\mA}_{j,j}\vy_r]_i\geq 0,i=1,2,...,m_o\}|$ and $|\{i\mid[\vy_o]_i\cdot[\hat{\mA}_{j,p'}\vy_r]_i\geq 0,i=1,2,...,m_o\}|,p'\in\cT_j-j$, conditioned on the same events as $E_{\Omega_j,n_j}$ is, as $B_j\sim B(m_o,p_j),p_j\geq1-\frac{\mbox{sin}^{-1}(1-\frac{\sqrt{2}(\|[\vx]_{\Omega\setminus 1}\|_2-\|[\vx]_{\Omega\setminus\Omega_j}\|_2)}{\|\vx\|_2})}{\pi}$ and $B_{j,p'}\sim B(m_o,p_{j,x}),p_{j,x}\leq1-\frac{\mbox{cos}^{-1}(\frac{\|[\vx]_{\{\Omega_{j-1}\cup j+1\}}\|_2}{\|\vx\|_2})}{\pi}$, respectively.


Now, we can compute the probability
\begin{align}
    \mathbb{P}\{E_{\Omega_j,n_j}\}= & \mathbb{P}\{\bigcap_{p'\in\cT_j-j}B_{j,p'}\leq n_j-1 \cap B_j\geq n_j\}\nonumber\\
    = & 1-\mathbb{P}\{\bigcup_{p'\in\cT_j-j}B_{j,p'}\geq n_j\cup B_j\leq n_j-1\}\nonumber\\
    \geq & 1-I_{\theta_{j,1}}(m_o-n_j+1,n_j)-(\kappa_j-1)\nonumber\\
    \times & [1-I_{\theta_{j,2}}(m_o-n_j+1,n_j)],
    \label{eq:Omega_j_n_j}
\end{align}
where $\theta_{j,1}=\frac{\mbox{sin}^{-1}(1-\frac{\sqrt{2}(\|[\vx]_{\Omega\setminus 1}\|_2-\|[\vx]_{\Omega\setminus\Omega_j}\|_2)}{\|\vx\|_2})}{\pi}$ and $\theta_{j,2}=\frac{\mbox{cos}^{-1}(\frac{\|[\vx]_{\{\Omega_{j-1}\cup j+1\}}\|_2}{\|\vx\|_2})}{\pi}$.

Finally, by combining (\ref{eq:succ}), (\ref{eq:cond}), (\ref{eq:j1}), (\ref{eq:jn}), (\ref{eq:cand1}), (\ref{eq:cond_LB1}), (\ref{eq:Ej_cj}), (\ref{eq:supp1_LB}), (\ref{eq:supp_a'_LB}), and (\ref{eq:Omega_j_n_j}), we obtain that
\begin{align}
    \mathbb{P}\{E_{\footnotesize{\mbox{succ}}}\}\geq & \prod_{j=2}^s\mathbb{P}\{E_{j,c_j}\mid E_{\varnothing \rightarrow j-1}\}\times\prod_{j=1}^s\mathbb{P}\{E_{\Omega_j,n_j}\}\nonumber\\
    \geq & \prod_{j=2}^s(1-\sqrt{\frac{2}{\pi}}c_j) [1-I_{e^{-\frac{{c_j}^2}{2}}}(\kappa_j,n-j+1-\kappa_j)]\nonumber\\
    \times & \prod_{j=1}^s\{ 1-I_{\theta_{j,1}}(m_o-n_j+1,n_j)-(\kappa_j-1)\nonumber\\
    \times & [1-I_{\theta_{j,2}}(m_o-n_j+1,n_j)]\}.
\end{align}

\section{Proof of Theorem 6}
\label{pf:Thm6}
Without loss of generality, assume that the first $s$ entries of $\vx$ are nonzero, while the remaining $n-s$ entries are equal to zero. Therefore, we denote the support set and the non-support set of $\vx$ as $\Omega=\{1,2,...,s\}$ and $\cN\cS=\{s+1,s+2,...,n\}$, respectively.
Also, assume that the initial set of detected support indexes is $\tilde{\Omega}_0=\{s'+1,s'+2,...,s\}$, \emph{i.e.}, the initial set of the $s'$ undetected support indexes is $\{1,2,...s'-1,s'\}$ with $|[\vx]_1|\geq|[\vx]_2|\geq...\geq|[\vx]_{s'-1}|\geq|[\vx]_{s'}|$,
and denote $\tilde{\Omega}_j$ as the set of detected indexes after the first $j$ iterations.

Now, we define two events, $E_{\footnotesize{\mbox{succ}}}$: $\tilde{\Omega}_{s'}=\Omega$ and $E_{\varnothing\rightarrow a'}$: the order of newly detected indexes after the first $a'$ iterations is $1\rightarrow 2\rightarrow ... \rightarrow a'$.
Consider a specific event $E_{\varnothing \rightarrow s'}$ where the order of newly detected indexes after the $s'$ iterations is: $1\rightarrow 2\rightarrow ... \rightarrow s'$. Note that when the event $E_{\varnothing \rightarrow s'}$ happens, the event $E_{\footnotesize{\mbox{succ}}}$ also happens. Therefore, it can be inferred that
\begin{equation}
    \mathbb{P}\{E_{\footnotesize{\mbox{succ}}}\}\geq\mathbb{P}\{E_{\varnothing \rightarrow s'}\}.
    \label{eq:succ2}
\end{equation}
With the multiplication law of probability, $\mathbb{P}\{E_{\varnothing \rightarrow s'}\}$ can be decomposed as
\begin{align}
    \mathbb{P}\{E_{\varnothing \rightarrow s'}\} & =\mathbb{P}\{E_{\varnothing \rightarrow 1}\}\times\mathbb{P}\{E_{1 \rightarrow 2}\mid E_{\varnothing \rightarrow 1}\}\times...\nonumber\\
    & \times\mathbb{P}\{E_{j-1 \rightarrow j}\mid E_{\varnothing \rightarrow j-1}\}\times...\nonumber\\
    & \times\mathbb{P}\{E_{s'-1 \rightarrow s'}\mid E_{\varnothing \rightarrow s'-1}\},
    \label{eq:Thm_7_cond}
\end{align}
where the event $E_{a'-1\rightarrow a'}$ is defined as: the newly detected index in the $(a'-1)$th iteration is $a'-1$ and the newly detected index in the $a'$th iteration is $a'$.

Define two events, $E_{\footnotesize{\mbox{aug}},a'}$: the index $a'$ is selected for support augmentation in the $a'$th iteration and $E_{\footnotesize{\mbox{pru}},a'}$: $\{\tilde{\Omega}_{a'-1}\cup a'\}\subseteq\tilde{\Omega}_{a'}$. Then, with the multiplication law of probability, the first term and the $a'$th term, $a'=2,3,...,s$, on the RHS of (\ref{eq:Thm_7_cond}) can be decomposed as
\begin{align}
    \mathbb{P}\{E_{\varnothing \rightarrow 1}\}= & \mathbb{P}\{E_{\footnotesize{\mbox{aug}},1}\cap E_{\footnotesize{\mbox{pru}},1}\}\nonumber\\
    = & \mathbb{P}\{E_{\footnotesize{\mbox{aug}},1}\}\times\mathbb{P}\{E_{\footnotesize{\mbox{pru}},1}\mid E_{\footnotesize{\mbox{aug}},1}\},
    \label{eq:Thm_7_j1}
\end{align}
\begin{align}
     \mathbb{P}\{E_{a'-1 \rightarrow a'}\mid E_{\varnothing \rightarrow a'-1}\} & =\mathbb{P}\{E_{\footnotesize{\mbox{aug}},a'}\cap E_{\footnotesize{\mbox{pru}},a'}\mid E_{\varnothing \rightarrow a'-1}\}\nonumber\\
     & =\mathbb{P}\{E_{\footnotesize{\mbox{aug}},a'}\mid E_{\varnothing \rightarrow a'-1}\}\nonumber\\
     & \times\mathbb{P}\{E_{\footnotesize{\mbox{pru}},a'}\mid E_{\footnotesize{\mbox{aug}},a'},E_{\varnothing \rightarrow a'-1}\},
     \label{eq:Thm_7_jn}
\end{align}
respectively.
In the following proof, we will address the probability terms on the RHS of (\ref{eq:Thm_7_j1}) and (\ref{eq:Thm_7_jn}).

To begin with, consider a specific index set $\Omega'\subset\{1,2,...,n\}$. For an analysis of the probability of the hyperplanes generated from $\mA_o$ separating $\vx$ and the estimate obtained from $\Omega'$, $\tau_n(\mA_{r_{\Omega'}}^{\dagger}\vy_r,\Omega')$, we would like to compute the $\ell_2$-norm between both points as
$\|\vx-\tau_n(\mA_{r_{\Omega'}}^{\dagger}\vy_r,\Omega')\|_2$.

In a similar manner to the proof of Theorem 5 (see Appendix \ref{pf:Thm5}), with the RIP of the matrix $\mA_r$ and $\|\ve_r\|_2\leq\frac{\|\vx\|_2-\sqrt{2}\|[\vx]_{\Omega-\{\tilde{\Omega}_0\cup 1\}}\|_2}{2+\sqrt{2}}$, it can be derived that
\begin{align}
    & \|\vx-\tau_n(\mA_{r_{\Omega'}}^{\dagger}\vy_r,\Omega')\|_2\nonumber\\
    \leq & \|\vx\|_2-\sqrt{2}(\|[\vx]_{\Omega\setminus\{\tilde{\Omega}_0\cup 1\}}\|_2-\|[\vx]_{\Omega\setminus\Omega'}\|_2).
\end{align}
Note that $\|\vx\|_2-\sqrt{2}(\|[\vx]_{\Omega\setminus\{\tilde{\Omega}_0\cup 1\}}\|_2-\|[\vx]_{\Omega\setminus\Omega'}\|_2)\leq\|\vx\|_2$ if $\{\tilde{\Omega}_0\cup 1\}\subset\Omega'$.

Now, consider the step of support augmentation. In the $j$th iteration, we would like to select the index $j$ from $\Omega_{j-1}^c$ and obtain the augmented estimated support set $\cT_j$.

Define two events, $E_{\tilde{\Omega}_1,\hat{n}_1}$: when $j=1$, $|\{i\mid[\vy_o]_i\cdot[\hat{\mA}_{1,1}\vy_r]_i\geq 0,i=1,2,...,m_o\}|\geq \hat{n}_1$ and $|\{i\mid[\vy_o]_i\cdot[\hat{\mA}_{1,p'}\vy_r]_i\geq 0,i=1,2,...,m_o\}|\leq \hat{n}_1-1,p'\in\{\Omega_0 \cup 1\}^c$, and $E_{\tilde{\Omega}_{a'},\hat{n}_{a'}}$: when $j=a',a'=2,3,...,s'$, $|\{i\mid[\vy_o]_i\cdot[\hat{\mA}_{a',a'}\vy_r]_i\geq 0,i=1,2,...,m_o\}|\geq \hat{n}_{a'}$ and $|\{i\mid[\vy_o]_i\cdot[\hat{\mA}_{a',p'}\vy_r]_i\geq 0,i=1,2,...,m_o\}|\leq \hat{n}_{a'}-1,p'\in\{\Omega_{a'-1}\cup a'\}^c$, conditioned on the event $E_{\varnothing \rightarrow a'-1}$, where $\hat{n}_j,j=1,2,...,s$ is the threshold number of support augmentation in the $j$th iteration.

Note that when the event $E_{\tilde{\Omega}_1,\hat{n}_1}$ happens, the event $E_{\footnotesize{\mbox{aug}},1}$ also happens, and when the event $E_{\tilde{\Omega}_{a'},\hat{n}_{a'}}$ happens, the event $E_{\footnotesize{\mbox{aug}},a'}$ conditioned on the event $E_{\varnothing \rightarrow a'-1}$ also happens. Therefore, it can be inferred that 
\begin{equation}
    \mathbb{P}\{E_{\footnotesize{\mbox{aug}},1}\}\geq\mathbb{P}\{E_{\tilde{\Omega}_1,\hat{n}_1}\},
    \label{eq:Thm_7_supp1_LB}
\end{equation}
\begin{equation}
    \mathbb{P}\{E_{\footnotesize{\mbox{aug}},a'}\mid E_{\varnothing \rightarrow a'-1}\}\geq\mathbb{P}\{E_{\tilde{\Omega}_{a'},\hat{n}_{a'}}\},
    \label{eq:Thm_7_supp_a'_LB}
\end{equation}
where the event $E_{\tilde{\Omega}_j,\hat{n}_j}$ implies that the set of detected support indexes in the $j$th iteration would be $\tilde{\Omega}_j=\{\tilde{\Omega}_0,1,2,...,j\}=\{\tilde{\Omega}_{j-1},j\}\subset\cT_j$ ($\tilde{\Omega}_0=\{s'+1,s'+2,...,s\}$).

It can be observed that the event $E_{\tilde{\Omega}_j,\hat{n}_j}$ in this step is similar to the event $E_{\Omega_j,n_j}$ in the step of support detection in the proof of Theorem 5 (see Appendix \ref{pf:Thm5}).
By replacing $n_j$, $\kappa_j$, $(\|[\vx]_{\Omega\setminus 1}\|_2-\|[\vx]_{\Omega\setminus\Omega_j}\|_2)$ in $\theta_{j,1}$, and $\Omega_{j-1}$ in $\theta_{j,2}$ on the RHS of (\ref{eq:Omega_j_n_j}) with $\hat{n}_j$, $n-s$, $(\|[\vx]_{\Omega\setminus\{\tilde{\Omega}_0\cup 1\}}\|_2-\|[\vx]_{\Omega\setminus\tilde{\Omega}_j}\|_2)$, and $\tilde{\Omega}_{j-1}$, respectively, we obtain that
\begin{align}
    \mathbb{P}\{E_{\tilde{\Omega}_j,\hat{n}_j}\}\geq & 1-I_{\hat{\theta}_{j,1}}(m_o-\hat{n}_j+1,\hat{n}_j)-(n-s-1)\nonumber\\
    \times & [1-I_{\hat{\theta}_{j,2}}(m_o-\hat{n}_j+1,\hat{n}_j)],
    \label{eq:Omega_j_hat_n_j}
\end{align}
where $\hat{\theta}_{j,1}=\frac{\mbox{sin}^{-1}(1-\frac{\sqrt{2}(\|[\vx]_{\Omega\setminus\{\tilde{\Omega}_0\cup 1\}}\|_2-\|[\vx]_{\Omega\setminus\tilde{\Omega}_j}\|_2)}{\|\vx\|_2})}{\pi}$ and $\hat{\theta}_{j,2}=\frac{\mbox{cos}^{-1}(\frac{\|[\vx]_{\{\tilde{\Omega}_{j-1}\cup j+1\}}\|_2}{\|\vx\|_2})}{\pi}$.

Next, consider the step of support pruning.
In the $j$th iteration, we would like to obtain a subset $\Omega_j$, which contains $\tilde{\Omega}_j$ and whose cardinality is $|\Omega_j|=s$, of the augmented estimated support set $\cT_j$, conditioned on the event $E_{\footnotesize{\mbox{aug}},j}$.
Since $|\cT_j|=s+1$, there are $\binom{s+1}{s}=s+1$ candidates for the selection of $\Omega_j$.
Among them, $s-s'+j$ candidates contain the set $\tilde{\Omega}_j$, and we would like to select any one of these candidates. The remaining $(s+1)-(s-s'+j)=s'-j+1$ candidates are not desired for the selection.

Define two events, $E_{\Omega_1,\tilde{n}_1}$: when $j=1$, $|\{i\mid[\vy_o]_i\cdot[\hat{\mA}_{\cT}\vy_r]_i\geq 0,i=1,2,...,m_o\}|\geq \tilde{n}_1,\tilde{\Omega}_1\subseteq\cT\subset\cT_1,|\cT|=s$ and $|\{i\mid[\vy_o]_i\cdot[\hat{\mA}_{\cT'}\vy_r]_i\geq 0,i=1,2,...,m_o\}|\leq \tilde{n}_1-1,\tilde{\Omega}_1\not\subseteq\cT',\cT'\subset\cT_1,|\cT'|=s$, conditioned on the event $E_{\footnotesize{\mbox{aug}},1}$, and $E_{\Omega_{a'},\tilde{n}_{a'}}$: when $j=a',a'=2,3,...,s'$, $|\{i\mid[\vy_o]_i\cdot[\hat{\mA}_{\cT}\vy_r]_i\geq 0,i=1,2,...,m_o\}|\geq \tilde{n}_{a'},\tilde{\Omega}_{a'}\subseteq\cT\subset\cT_{a'},|\cT|=s$ and $|\{i\mid[\vy_o]_i\cdot[\hat{\mA}_{\cT'}\vy_r]_i\geq 0,i=1,2,...,m_o\}|\leq \tilde{n}_{a'}-1,\tilde{\Omega}_{a'}\not\subseteq\cT',\cT'\subset\cT_{a'},|\cT'|=s$, conditioned on the events $E_{\footnotesize{\mbox{aug}},a'}$ and $E_{\varnothing \rightarrow a'-1}$, where $\tilde{n}_j,j=1,2,...,s$ is the threshold number of support pruning in the $j$th iteration.

Note that when the event $E_{\Omega_1,\tilde{n}_1}$ happens, the event $E_{\footnotesize{\mbox{pru}},1}$ conditioned on the event $E_{\footnotesize{\mbox{aug}},1}$ also happens, and when the event $E_{\Omega_{a'},\tilde{n}_{a'}}$ happens, the event $E_{\footnotesize{\mbox{pru}},a'}$ conditioned on the events $E_{\footnotesize{\mbox{aug}},a'}$ and $E_{\varnothing \rightarrow a'-1}$ also happens. Therefore, it can be inferred that 
\begin{equation}
    \mathbb{P}\{E_{\footnotesize{\mbox{pru}},1}\mid E_{\footnotesize{\mbox{aug}},1}\}\geq\mathbb{P}\{E_{\Omega_1,\tilde{n}_1}\},
    \label{eq:Thm_7_pru1_LB}
\end{equation}
\begin{equation}
    \mathbb{P}\{E_{\footnotesize{\mbox{pru}},a'}\mid E_{\footnotesize{\mbox{aug}},a'},E_{\varnothing \rightarrow a'-1}\}\geq\mathbb{P}\{E_{\Omega_{a'},\tilde{n}_{a'}}\},
    \label{eq:Thm_7_pru_a'_LB}
\end{equation}
where the event $E_{\Omega_j,\tilde{n}_j}$ implies that the pruned estimated support set $\Omega_j\supset\tilde{\Omega}_j$.

It can be observed that the event $E_{\Omega_j,\tilde{n}_j}$ in this step is also similar to the event $E_{\Omega_j,n_j}$ in the step of support detection in the proof of Theorem 5 (see Appendix \ref{pf:Thm5}).
By replacing $n_j$, $(\kappa_j-1)$, $(\|[\vx]_{\Omega\setminus 1}\|_2-\|[\vx]_{\Omega\setminus\Omega_j}\|_2)$ in $\theta_{j,1}$, and $\{\Omega_{j-1}\cup j+1\}$ in $\theta_{j,2}$ on the RHS of (\ref{eq:Omega_j_n_j}) with $\tilde{n}_j$, $s'-j+1$, $(\|[\vx]_{\Omega\setminus\{\tilde{\Omega}_0\cup 1\}}\|_2-\|[\vx]_{\Omega\setminus\tilde{\Omega}_j}\|_2)$, and $\tilde{\Omega}_j\setminus i_j^*$ (with $i_j^*=\underset{i\in\tilde{\Omega}_j}{\mbox{argmin}}|[\vx]_i|$), respectively, and multiplying $s-s'+j$ to the second term on the RHS of (\ref{eq:Omega_j_n_j}), we obtain that
\begin{align}
    \mathbb{P}\{E_{\Omega_j,\tilde{n}_j}\}\geq & 1-(s-s'+j)\times I_{\hat{\theta}_{j,1}}(m_o-\tilde{n}_j+1,\tilde{n}_j)\nonumber\\
    - & (s'-j+1)[1-I_{\tilde{\theta}_{j,2}}(m_o-\tilde{n}_j+1,\tilde{n}_j)],
    \label{eq:Omega_j_tilde_n_j}
\end{align}
where $\tilde{\theta}_{j,2}=\frac{\mbox{cos}^{-1}(\frac{\|[\vx]_{\tilde{\Omega}_j\setminus i_j^*}\|_2}{\|\vx\|_2})}{\pi}$.

Finally, by combining (\ref{eq:succ2}), (\ref{eq:Thm_7_cond}), (\ref{eq:Thm_7_j1}), (\ref{eq:Thm_7_jn}), (\ref{eq:Thm_7_supp1_LB}), (\ref{eq:Thm_7_supp_a'_LB}), (\ref{eq:Omega_j_hat_n_j}), (\ref{eq:Thm_7_pru1_LB}), (\ref{eq:Thm_7_pru_a'_LB}), and (\ref{eq:Omega_j_tilde_n_j}), we derive that
\begin{align}
    \mathbb{P}\{E_{\footnotesize{\mbox{succ}}}\}\geq\prod_{j=1}^s & \mathbb{P}\{E_{\tilde{\Omega}_j,\hat{n}_j}\}\times\prod_{j=1}^s\mathbb{P}\{E_{\Omega_j,\tilde{n}_j}\}\nonumber\\
    \geq\prod_{j=1}^s & \{1-I_{\hat{\theta}_{j,1}}(m_o-\hat{n}_j+1,\hat{n}_j)-(n-s-1)\times\nonumber\\
    & [1-I_{\hat{\theta}_{j,2}}(m_o-\hat{n}_j+1,\hat{n}_j)]\}\nonumber\\
    \times & \{1-(s-s'+j)\times I_{\tilde{\theta}_{j,1}}(m_o-\tilde{n}_j+1,\tilde{n}_j)\nonumber\\
    - & (s'-j+1)[1-I_{\tilde{\theta}_{j,2}}(m_o-\tilde{n}_j+1,\tilde{n}_j)]\}.
\end{align}

\end{appendices}
\ifCLASSOPTIONcaptionsoff
  \newpage
\fi



%


\bibliographystyle{IEEEtran}
\bibliography{IEEEabrv,waveform}

%





\end{document}